\documentclass[12pt]{article}

 \usepackage{epsfig}
 \usepackage{pstricks}
 \usepackage{cite}

\def\ve{\varepsilon}

\def\m{\mu}
\def\beq{\begin{equation}}
\def\eeq{\end{equation}}
\def\b{\beta}
\def\beqa{\begin{eqnarray}}
\def\eeqa{\end{eqnarray}}
\def\D{\Delta}
\def\G{\Gamma}

\def\n{\nu}
\def\p{\pi}
\def\k{\kappa}
\def\cl{{\cal L}}
\def\Bar#1{\overline{#1}}                       
\def\VEV#1{\left\langle #1\right\rangle}        
\def\dg{\dagger}

\def\NO{\nonumber}
\def\mt{\widetilde{m}_1}                  
\def\mb{\overline{m}}
\def\pl#1#2#3{Phys.~Lett.~{\bf B{#1}}, #3 ({#2})}
\def\np#1#2#3{Nucl.~Phys.~{\bf B{#1}}, #3 ({#2})}
\def\prl#1#2#3{Phys.~Rev.~Lett.~{\bf #1}, #3 ({#2})}
\def\pr#1#2#3{Phys.~Rev.~{\bf D{#1}}, #3 ({#2})}

\begin{document}
\title{
\vspace*{-2cm}
{\normalsize
\begin{minipage}{3cm}
DESY 05-031\\
February 2005
\end{minipage}}\hspace{\fill}\mbox{}\\[5ex]
{\bf\large LEPTOGENESIS AS THE ORIGIN OF MATTER}
}
\author{W. Buchm\"uller$^a$, R. D. Peccei$^b$, T. Yanagida$^c$  \\
\vspace{3.0\baselineskip}                                               
{\normalsize\it a Deutsches Elektronen-Synchrotron DESY, 22603 Hamburg, 
Germany}
\\
\vspace{3.0\baselineskip}                                               
{\normalsize\it b Department of Physics and Astronomy, University of California
at Los Angeles,}\\ {\normalsize\it Los Angeles, California, 90095, USA}\\
\vspace{3.0\baselineskip}                                               
{\normalsize\it c Department of Physics, University of Tokyo, Tokyo 113-0033, 
Japan}\\
}

\date{}
\maketitle
\thispagestyle{empty}
\begin{abstract}
\noindent
We explore in some detail the hypothesis that the generation of a
primordial lepton-antilepton asymmetry (Leptogenesis) early on in the
history of the Universe is the root cause for the origin of matter. After
explaining the theoretical conditions for producing a matter-antimatter
asymmetry in the Universe we detail how, through sphaleron processes, it
is possible to transmute a lepton asymmetry -- or, more precisely, a
(B-L)-asymmetry -- into a baryon asymmetry. Because Leptogenesis
depends in detail on properties of the neutrino spectrum, we review
briefly existing experimental information on neutrinos as well as the
seesaw mechanism, which offers a theoretical understanding of why
neutrinos are so light. The bulk of the review is devoted to a discussion
of thermal Leptogenesis and we show that for the neutrino spectrum
suggested by oscillation experiments one obtains the observed value for
the baryon to photon density ratio in the Universe, independently of any
initial boundary conditions. In the latter part of the review we consider
how well Leptogenesis fits with particle physics models of dark matter.
Although axionic dark matter and Leptogenesis can be very naturally
linked, there is a potential clash between Leptogenesis and models of
supersymmetric dark matter because the high temperature needed for
Leptogenesis leads to an overproduction of gravitinos, which alter the
standard predictions of Big Bang Nucleosynthesis. This problem can be
resolved, but it constrains the supersymmetric spectrum at low energies
and the nature of the lightest supersymmetric particle (LSP). Finally, as an 
illustration of possible other
options for the origin of matter, we discuss the possibility that
Leptogenesis may occur as a result of non-thermal processes.
\end{abstract}

\maketitle

\newpage
\tableofcontents


\section{Introduction}

Our understanding of the Universe has deepened considerably in the last
25 years, so much so that a standard cosmological model has emerged
\cite{SCM}. In this model, after the Big Bang, a period of inflationary
expansion \cite{inflation} ensued that effectively set the Universe's
curvature to zero. After inflation the Universe's expansion continued,
not in an exponential fashion but with the rate of expansion being
determined by which component of the Universe's energy density dominated
the total energy density.

In the present epoch this energy density is dominated by a, so-called,
dark energy component whose negative pressure causes the Universe's
expansion to accelerate \cite{DE}. Dark energy now accounts for
approximately 70\% of the total energy density, with the other 30\% of
the remaining energy density of the Universe's being dominated by some
kind of non-luminous (dark) matter. In detail, the angular distribution
of the temperature fluctuations of the microwave background radiation
measured by the 
Wilkinson Microwave Anisotropy Probe (WMAP)
collaboration \cite{WMAP} determines the various
components of the ratio of the Universe's energy density now $\rho_o$ to
the critical energy density $\rho_c$, 
$\Omega= \rho_o/\rho_c$.\footnote{The critical 
density $\rho_c$ is the density that
corresponds to a closed Universe now, $ \rho_c= 3H_o^2/8\pi G_N$. Here
$H_o$ is the value of the Hubble parameter now. Inflation predicts that $\rho_o=\rho_c$, so that
$\Omega=1$. } The results are: $\Omega_{\rm{dark~energy}}= 0.73\pm 0.04$;
$\Omega_{\rm{matter}}= 0.27 \pm 0.04$; and $\Omega_{\rm{B}}= 0.044 \pm
0.004$. Here $\Omega_{\rm{B}}$ is the contribution of baryonic matter to
$\Omega_{\rm{matter}}$, confirming that about 85 \% of $\Omega_{\rm{matter}}$ is indeed contributed by dark matter. The contribution
of neutrinos and photons is at a few per mil, or below, and is negligible.

Although the standard cosmological model sketched above provides an
accurate description of the present Universe and its evolution, deep
questions remain to be answered. What exactly constitutes the dark
energy? Is it just a cosmological constant? But if that is so, why is
the energy scale associated with the corresponding vacuum energy density
[$\rho_{\rm{cc}}=E_o^4; E_o \simeq 2 \times 10^{-3}$ eV] so small?
Equally mysterious is the nature of the dark matter, although in this
case there are at least some particle physics candidates that may be the
source for this component of the Universe's energy density.

A further mystery is associated with the observed baryon energy density.
This number can be used to infer the ratio of the number density of
baryons to photons in the Universe, a quantity that is measured
independently from the primordial nucleosynthesis of light elements.
The WMAP results \cite{WMAP} are in agreement with the most recent
nucleosynthesis analysis of the primordial Deuterium abundance, but there 
are discrepancies with both the inferred $\rm{^4He}$ and $\rm{^7Li}$ 
values \cite{nucleo}. These latter values, however, may have an 
underestimated error \cite{OS}. Averaging the WMAP  result only with that 
coming from the primordial abundance of Deuterium gives:
\begin{equation}
\frac{n_B}{n_{\gamma}}\equiv \eta_B= 6.1 \pm 0.3 \times
10^{-10}.
\end{equation}
Why does this ratio have this value?

In this review, we will principally try to address this last question
which, as we shall see, is intimately related to the existence of a
primordial matter-antimatter asymmetry. Nevertheless, 
we shall try, when germane, to connect our discussion with the broader
issues of what constitutes dark energy and dark matter. 

There is good evidence that the Universe is mostly made up of matter, 
although it is possible that small amounts of antimatter exist
\cite{anti}. However, antimatter certainly does not constitute one of the
dominant components of the Universe's energy density. Indeed, as Cohen,
de Rujula, and Glashow \cite{CDG} have compellingly argued, if there
were to exist large areas of antimatter in the Universe they could only
be at a cosmic distance scale from us. Thus, along with the question of
why $n_B/n_{\gamma}$ has the value given in Eq. (1), there is a parallel
question of why the Universe is predominantly composed of baryons rather
than antibaryons.


In fact, these two questions are interrelated. If the Universe had been
matter-antimatter symmetric at temperatures of O(1 GeV), as the
Universe cools further and the inverse process $ 2\gamma \to B+ \bar{B}$
becomes ineffective because of the Boltzmann factor, the number density
of baryons and antibaryons relative to photons would have been reduced
dramatically as a result of  the annihilation process $ B+ \bar{B} \to
2\gamma $.   A straightforward calculation gives, in this case,
\cite{ann}:
\begin{equation}
\frac{n_B}{n_{\gamma}}=\frac{n_{\bar{B}}}{n_{\gamma}} \simeq 10^{-18}.
\end{equation}
Thus, in a symmetric Universe the question is really why
observationally $n_B/n_{\gamma}$ is so large!

It is very difficult to imagine processes at temperatures below a GeV
that could enhance the ratio of the number density of baryons relative
to that of photons much beyond the value this quantity attains when
baryon-antibaryon annihilation occurs.\footnote{ An exception is provided by some versions of Affleck-Dine Baryogenesis \cite{AD} where a baryon excess is produced by the decay of a scalar field very late in the history of the Universe, which reheats the Universe to temperatures of the 
${\cal O}(100\ {\rm MeV})$.} 
Thus, because Eq. (2) does not
agree with the observed value given in Eq. (1), one is led to the
interesting conclusion that a primordial matter-antimatter asymmetry
must have existed at temperatures of ${\cal O}(1\ {\rm GeV})$ in the 
Universe. The
observed value for $n_B/n_{\gamma}$ and the lack of antimatter in the
Universe are manifestations of this primordial asymmetry. Hence, in
reality, the ratio $\eta_B$ is, in effect, a measure of the number density
of matter minus that of antimatter relative to the photon number
density:
\begin{equation}
\eta_B=\frac{n_B- n_{\bar{B}}}{n_{\gamma}}= 6.1 \pm 0.3 \times
10^{-10}.
\end{equation}

It is interesting to consider the physical origins of
this primordial matter-antimatter asymmetry. From the seminal work of
Sakharov \cite{Sakharov} one knows that, under certain conditions which
we will amplify later on in this article, this asymmetry can be
generated by physical processes. In this review we will focus on
Leptogenesis -- the creation of a primordial lepton-antilepton
asymmetry -- as the root source for the observed baryon- antibaryon
asymmetry of Eq. (3) \cite{FY}. In our view, Leptogenesis provides the most
compelling scenario for generating the observed baryon asymmetry in the
Universe. In particular, because Leptogenesis is closely linked with
parameters in the neutrino sector that can be eventually determined
experimentally, this scenario can be tested and can be either confirmed
or ruled out by data.

\enlargethispage{0.5cm} 

The plan of this review is as follows. In Section 2 we discuss the
theoretical conditions necessary for producing a primordial matter-
antimatter asymmetry in the Universe and explain how, through a
mechanism first discussed by Kuzmin, Rubakov, and Shaposhnikov,
\cite{KRS}, it is possible to turn a lepton asymmetry into a baryon
asymmetry. In Section 3 we review existing experimental information on
the neutrino sector, as well as the seesaw mechanism \cite{seesaw} that
provides a theoretical framework for understanding this data. Section 4
discusses thermal Leptogenesis and contains the main quantitative
results of the review. In particular, we show in this Section that the
observed value for $\eta_B$ obtained from Leptogenesis significantly
constrains low energy neutrino properties, and vice versa.  In Section 5
we turn to dark matter and discuss how supersymmetric candidates for
dark matter are significantly constrained if thermal Leptogenesis is the source
of the observed baryon asymmetry in the Universe. The constraint arises
from the overproduction of gravitinos. Section 6
discusses nonthermal Leptogenesis and other nonthermal processes that
can lead to Baryogenesis. Finally, we present our conclusion and summary
of results in Section 7.

\section{Theoretical Foundations}
\vspace{-0.2cm}
\subsection{Sakharov's Conditions for Baryogenesis}
\vspace{-0.1cm}
In 1967, Sakharov \cite{Sakharov} considered the consequences of the hypothesis that the observed
expanding Universe originated from a superdense initial state with temperature
of order the Planck mass, $T_i \sim M_{\rm P}$. Since he could not imagine
how, starting from such an initial state, one could obtain a macroscopic
separation of matter and antimatter, he concluded that our Universe contains
today only matter. That is, the Universe evolved from an initial state even under 
charge conjugation to a state odd under charge conjugation today. He then
realized that in an expanding Universe a matter-antimatter asymmetry could 
be generated dynamically, if C, CP, and baryon and lepton number were violated, and these processes were out of thermal equilibrium.

Sakharov also described a concrete model for Baryogenesis. He proposed as the 
origin for
the baryon and lepton asymmetry the CP-violating decays of maximons,  hypothetical
neutral spin zero particles with mass of order the Planck mass. Their
existence leads to a departure from thermal equilibrium already  at 
temperatures $T \sim M_{\rm P}$, where a small matter-antimatter asymmetry
is then generated. An unavoidable consequence of this model is that protons are unstable and
decay. However, the proton lifetime in Sakharov's model turned out to be unobservably long,
$\tau_p > 10^{50}\ {\rm years}$.

During the past four decades many models of Baryogenesis have been proposed,
demonstrating that the conditions Sakharov spelled out to allow Baryogenesis to take place are quite readily satisfied
in the Standard Model of particle physics and its extensions. There are,
however, significant differences among the various mechanisms suggested for producing  the
baryon asymmetry. Grand Unified Theories (GUTs) have been of particular importance
for the development of realistic models of Baryogenesis \cite{yo78}. These theories
provide natural heavy particle candidates, whose decays can be the source of the baryon
asymmetry. However, in general, the simplest GUT models based on $SU(5)$ lead 
to a creation of a (B+L)-asymmetry, with a vanishing asymmetry for B-L. As 
will be made clear below, a (B+L)-asymmetry generated at the GUT scale 
eventually gets erased by sphaleron processes. In Leptogenesis, heavy Majorana 
neutrinos required by the seesaw
mechanism \cite{seesaw} serve to trigger Baryogenesis. Because B-L is violated, the erasure present in GUTs is avoided. In principle, Electroweak Baryogenesis \cite{electro}
is also an attractive possibility, as the relevant parameters could then be tested in
collider experiments. However, in general, the electroweak phase transition is not sufficiently out of equilibrium to generate an asymmetry of the magnitude observed in the Universe. Finally, in supersymmetric theories the baryon and lepton
number stored in scalar expectation values can also lead to Baryogenesis, through the, so-called,
Affleck-Dine mechanism \cite{AD}, which will be discussed in some detail in Section 6.

\enlargethispage{0.7cm} 

\newpage

\subsection{$B+L$ Violation in the Standard Model}

Due to the chiral nature of the electroweak interactions, baryon and lepton 
number are not conserved in the Standard Model \cite{tH}. The divergence of the
$B$ and $L$ currents,
\begin{eqnarray}
J^B_\mu &=& {1\over 3} \sum_{generations}\left(\overline{q_L}\gamma_\mu q_L
+ \overline{u_R}\gamma_\mu u_R + \overline{d_R}\gamma_\mu d_R\right)\;,\\ 
J^L_\mu &=& \sum_{generations}\left(\overline{l_L}\gamma_\mu l_L
+ \overline{e_R}\gamma_\mu e_R \right)\;, 
\end{eqnarray}
is given by the triangle anomaly, 
\begin{eqnarray}
\partial^\mu J^B_\mu &=& \partial^\mu J^L_\mu \nonumber\\ 
&=& {N_f\over 32 \pi^2} \left(-g^2 W^I_{\mu\nu}\widetilde{W}^{I\mu\nu}
+ g'^2 B_{\mu\nu}\widetilde{B}^{\mu\nu}\right)\;.
\end{eqnarray}
Here $N_f$ is the number of generations, and $W^I_{\mu}$ and $B_{\mu}$ are, 
respectively, the $SU(2)$ and $U(1)$ gauge fields with gauge couplings $g$ 
and $g'$.

As a consequence of the anomaly, the change in baryon and lepton number is
related to the change in the topological charge of the gauge field,
\begin{eqnarray}
B(t_f) - B(t_i) &=& \int_{t_i}^{t_f}dt\int d^3x\partial^\mu J^B_\mu \nonumber\\
&=&N_f[ N_{cs}(t_f) - N_{cs}(t_i)]\;,\label{deltaB}
\end{eqnarray} 
where 
\begin{equation}
N_{cs}(t) = {g^3\over 96\pi^2} \int d^3x \epsilon_{ijk}\epsilon^{IJK}
W^{Ii}W^{Jj}W^{Kk}\;.
\end{equation}
For vacuum to vacuum transitions $W^{Ii}$ is a pure gauge configuration
and the Chern-Simons numbers $N_{cs}(t_i)$ and $N_{cs}(t_f)$ are integers.

In a non-abelian gauge theory there are infinitly many degenerate ground
states, which differ in their value of the Chern-Simons number,
$\Delta N_{cs} = \pm 1, \pm2, \ldots$.  The correponding points in field
space are separated by a potential barrier whose height is given by the
so-called sphaleron energy $E_{sph}$ \cite{Manton}. Because of the anomaly, 
jumps in the
Chern-Simons number are associated with changes of baryon and lepton number,
\begin{equation}
\Delta B = \Delta L = N_f \Delta N_{cs}\;.
\end{equation}
Obviously, in the Standard Model the smallest jump is $\Delta B = \Delta L = \pm 3$.

In the semiclassical approximation, the probability of tunneling between neighboring vacua is determined by instanton configurations. In the Standard Model, 
$SU(2)$ instantons lead to an
effective $12$-fermion interaction 
\begin{equation}\label{obl}
  O_{B+L} = \prod\limits_{i=1\ldots 3}\left(q_{Li}q_{Li}q_{Li}l_{Li}\right)\;,
\end{equation}
which describes processes with $\Delta B= \Delta L=3$, such as
\begin{equation}
u^c + d^c + c^c \rightarrow d + 2 s + 2 b +t + \nu_e + \nu_\mu + \nu_\tau\;.
\end{equation}
The transition rate is determined by the instanton action and one finds \cite{tH}
\begin{eqnarray}
\Gamma &\sim& e^{-S_{\rm inst}} = e^{-{4\pi\over \alpha}} \nonumber\\
&=& {\cal O}\left(10^{-165}\right)\;.
\end{eqnarray}
Because this rate is extremely small, ($B+L$)-violating interactions
appear to be completely negligible in the Standard Model. However, this picture changes dramatically when one is in a thermal bath. 

\subsection{Sphalerons and the KRS Mechanism}

As emphasized in the seminal paper of Kuzmin, Rubakov and Shaposhnikov 
\cite{KRS},
in the thermal bath provided by the expanding Universe one can make transitions between the gauge vacua not by tunneling, but through thermal fluctuations over the barrier. For temperatures larger than
the height of the barrier, the exponential suppresion in the rate provided by the Boltzmann factor disappears completely. Hence (B+L)-violating processes can occur at a significant rate and these processes can be in equilibrium in the expanding Universe.

The finite-temperature transition rate in the electroweak theory is
determined by the sphaleron configuration \cite{Manton}, a saddle point of the field energy
of the gauge-Higgs system. Fluctuations around this saddle point have
one negative eigenvalue, which allows one to extract the transition rate.
The sphaleron energy is proportional to $v_F(T)$, the finite-temperature 
expectation value of the Higgs field, and one finds
\begin{equation}
E_{sph}(T) \simeq {8\pi\over g} v_F(T)\;.
\end{equation}
Taking translational and rotational zero-modes into account, one obtains
for the transition rate per unit volume in the Higgs phase \cite{aml87}
\begin{equation}
{\Gamma_{B+L}\over V}=\kappa {M_W^7\over (\alpha T)^3} e^{-\beta E_{sph}(T)}\;,
\end{equation}
where $\beta = 1/T$, $M_W = g^2 v_F(T)/2$ and $\kappa$ is some constant.

Extrapolating this semiclassical formula to the high-temperature symmetric 
phase, where $v_F(T) = 0$, and using for $M_W$ the thermal mass, $M_W\sim g^2 T$,
one expects in this phase $\Gamma_{B+L}/V \sim (\alpha T)^4$.
However, detailed studies have shown that this naive extrapolation from the Higgs to
the symmetric phase is not quite correct. The relevant spatial scale for 
non-perturbative fluctuations is the magnetic screening length 
$\sim 1/(g^2 T)$, but the corresponding time
scale turns out to be $1/(g^4 T \ln{g^{-1}})$, which is larger for small
coupling \cite{yaffe,bodeker}. As a consequence one obtains for the sphaleron rate in the symmetric phase
\begin{equation}
\Gamma_{B+L}/V \sim \alpha^5 \ln{\alpha^{-1}} T^4.
\end{equation}

It turns out that the dynamics of low-frequency gauge fields can be described by a remarkably simple
effective theory, derived by B\"odeker \cite{bodeker}. The color magnetic and 
electric fields satisfy the equation of motion
\begin{equation}
\vec{D}\times\vec{B} = \sigma \vec{E} - \vec{\zeta}\;.
\end{equation}
Here $\vec{\zeta}$ is Gaussian noise, a random vector field with variance
\begin{equation}
\langle \zeta_i(x)\zeta_j(x')\rangle = 2\sigma \delta_{ij} \delta^4(x-x')\;.
\end{equation}
These equations define a stochastic three-dimensional gauge theory. The 
parameter $\sigma$ is the `color conductivity', $\sigma = m_D^2/(3\gamma)$, where $m_D \sim gT$ is the
Debye screening mass and $\gamma \sim g^2T \ln(1/g)$ is the hard gauge boson damping rate. To leading-log accuracy one has 
$1/\sigma \sim \ln{g^{-1}}$. A next-to-leading order analysis yields for the 
sphaleron rate \cite{yaffe2}
\begin{equation}\label{sphrate}
{\Gamma_{B+L}\over V} = (10.8\pm 0.7) \left({gT\over m_D}\right)^2 
\alpha^5 T^4 \left[\ln{\left({m_D\over \gamma}\right)} + 3.041
+ \left({1\over \ln{(1/g)}}\right)\right]\;.
\end{equation}
The overall coefficient has been determined by a numerical lattice 
simulation \cite{bmr00}. From Eq.~(\ref{sphrate}) one easily obtains the temperature range
where sphaleron processes are in thermal equilibrium:
\begin{equation}\label{range}
T_{EW} \sim 100\ {\rm GeV} < T < T_{sph} \sim 10^{12}\ {\rm GeV}\;.
\end{equation} 

The effective theory describing topological fluctuations of the gauge field
in the high-temperature phase is valid for small coupling, $g\ll 1$. Yet for
$T_{EW} < T < T_{sph} \sim 10^{12}\ {\rm GeV}$ one has $g={\cal O}(1)$. This implies
that the electric screening lenth $1/(gT)$ and the magnetic screening length 
$1/(g^2 T)$ are not well separated and that nonperturbative corrections
to the sphaleron rate, Eq. (\ref{sphrate}), may be large. This will modify the
temperature range given in Eq. (\ref{range}), but one expects that the qualitative picture
of fluctuations in baryon and lepton number in the high-temperature phase
of the Standard Model will not be affected.

\subsection{Electroweak Baryogenesis and its Experimental Constraints}

An important ingredient in the theory of Baryogenesis is related to the nature
of the electroweak transition from the high-temperature symmetric phase to the
low-temperature Higgs phase. Because in the Standard Model baryon number, C and
CP are not conserved, it is conceivable that the cosmological
baryon asymmetry could have been generated at the electroweak phase 
transition \cite{KRS}, provided that this transition is of first-order, 
because then there is also the necessary departure from thermal equilibrium. 
This possibility has stimulated much theoretical
activity during the past years to determine the phase diagram of the
electroweak theory. 

Electroweak Baryogenesis requires that the baryon asymmetry 
generated during the phase transition is not erased by sphaleron processes
afterwards.\footnote{ The produced asymmetry will be erased if, after the phase transition, (B+L)-violating processes are in equilibrium.} This leads to a condition on the jump of the Higgs vacuum
expectation value $v_F =\sqrt{H^{\dagger}H}$ at the critical 
temperature \cite{sh86}:
\begin{equation}\label{sphbound}
{\Delta v_F(T_c)\over T_c} > 1\;.
\end{equation}
The strength of the electroweak transition has been studied by numerical and
analytical methods as function of the Higgs boson mass. For the
$SU(2)$ gauge-Higgs model one finds from lattice simulations as well as
perturbative calculations that the lower bound of Eq. (\ref{sphbound}) is 
violated for Higgs masses above 45 GeV \cite{ja96}.\footnote{ For Higgs masses
below 50 GeV, the Higgs model provides a good approximation for the full Standard 
Model.} Because the present lower bound from LEP on the Higgs mass is
114 GeV \cite{PDG}, it is clear that the electroweak transition in the Standard Model is too weak for Baryogenesis. However, for special choices of 
parameters or by adding singlet fields, in certain circumstances supersymmetric extensions of the Standard
Model have a sufficiently strong first-order phase transition to allow 
Electroweak Baryogenesis to take place \cite{susyew}. 

For large Higgs masses, the nature of the electroweak transition is dominated 
by nonperturbative effects of the $SU(2)$ gauge theory at high temperatures. 
At a critical Higgs mass $m_H^c = {\cal O}(M_W)$, an intriguing phenomenon 
occurs: The first-order phase transition turns into a smooth 
crossover \cite{bp95,klrs96,bw97}, as expected on general grounds \cite{ja96}. 
At the endpoint of a critical line of first-order transitions, which is reached
for $m_H=m_H^c$, the phase transition is of second order \cite{rtk98}.

The value of the critical Higgs mass can be estimated by comparing the
W-boson mass $M_W$ in the Higgs phase with the magnetic mass $m_{SM}$ in the 
symmetric phase. This yields $m_H^c \simeq 74\ \mbox{GeV}$ \cite{bp97}.
Numerical lattice simulations
have determined the precise value $m_H^c = 72.1 \pm 1.4$~GeV \cite{fo99}. 
The analytic estimate of the critical Higgs mass can be generalized to  
supersymmetric extensions of the Standard Model, where one finds 
$m_h^c < 130\dots 150$~GeV \cite{ce97}, which is still compatible with
the present experimental lower bound.

\subsection{The Relation Between Baryon and Lepton Asymmetries}

In a weakly coupled plasma, one can assign a chemical potential $\m$ to 
each of the quark, lepton and Higgs fields. In the Standard Model, with one 
Higgs doublet $H$ and $N_f$ generations one then has $5N_f+1$ chemical 
potentials.\footnote{In addition to the Higgs doublet, the two left-handed
doublets $q_i$ and $\ell_i$ and the three right-handed singlets $u_i$, $d_i$,
and $e_i$ of each generation each have an independent chemical potential.}
For a non-interacting gas of massless particles the asymmetry in the particle 
and antiparticle number densities is given by
\begin{equation}\label{number}
n_i-\overline{n}_i={g T^3\over 6}
\left\{\begin{array}{rl}\b\m_i +{\cal O}\left(\left(\b\m_i\right)^3\right)\;,
&\mbox{fermions}\;,\\
2\b\m_i+{\cal O}\left(\left(\b\m_i\right)^3\right)\;, &\mbox{bosons}\;.
\end{array}\right.
\end{equation}
The following analysis is based on these relations for $\b \m_i \ll 1$.
However, one should keep in mind that the plasma of the early Universe is
very different from a weakly coupled relativistic gas, owing to the presence
of unscreened non-abelian gauge interactions, where nonperturbative effects
are important in some cases.

Quarks, leptons and Higgs bosons interact via
Yukawa and gauge couplings and, in addition, via the nonperturbative
sphaleron processes. In thermal equilibrium all these processes yield
constraints between the various chemical potentials \cite{ht}. The effective 
interaction
of Eq. (\ref{obl}) induced by the $SU(2)$ electroweak instantons implies 
\begin{equation}\label{sphew}
\sum_i\left(3\m_{qi} + \m_{li}\right) = 0\;.
\end{equation}
One also has to take the $SU(3)$ Quantum Chromodynamics (QCD)
instanton processes into account 
\cite{moh}, which generate an effective interaction
between left-handed and right-handed quarks. The corresponding relation
between the chemical potentials reads
\begin{equation}\label{sphqcd}
\sum_i\left(2\m_{qi} - \m_{ui} - \m_{di}\right) = 0\;.
\end{equation}
A third condition, valid at all temperatures, arises from the
requirement that the total hypercharge of the plasma vanishes. From
Eq.~(\ref{number}) and the known hypercharges one obtains
\begin{equation}\label{hypsm}
\sum_i\left(\m_{qi} + 2 \m_{ui} - \m_{di} - \m_{li} - \m_{ei} + 
{2\over N_f} \m_{H}\right) = 0\;.
\end{equation}

The Yukawa interactions, supplemented by gauge interactions, yield relations
between the chemical potentials of left-handed and right-handed fermions,
\begin{equation}\label{myuk}
\m_{qi}-\m_{H}-\m_{dj} = 0\;, \quad
\m_{qi}+\m_{H}-\m_{uj} = 0\;, \quad
\m_{li}-\m_{H}-\m_{ej} = 0\;.
\end{equation}
These relations hold if the corresponding interactions are in thermal 
equilibrium. In the temperature range
$100\ \mbox{GeV} < T < 10^{12}\ \mbox{GeV}$, 
which is of interest for Baryogenesis, this is the case for gauge 
interactions. On the other hand, Yukawa interactions are in equilibrium only
in a more restricted temperature range that depends on the strength of the 
Yukawa couplings. In the following we shall ignore this complication which
has only a small effect on our discussion of Leptogenesis.

Using Eq.~(\ref{number}), the baryon number density $n_B \equiv g B T^2/6$
and the lepton number densities $n_{L_i} \equiv L_i gT^2/6$ can be expressed in 
terms of the chemical potentials: 
\begin{eqnarray}
B &=& \sum_i \left(2\m_{qi} + \m_{ui} + \m_{di}\right)\;, \\
L_i &=& 2\m_{li} + \m_{ei}\;,\quad L=\sum_i L_i\;.
\end{eqnarray}

Consider now the case where all Yukawa interactions are in equilibrium. 
The asymmetries $L_i-B/N_f$ are then conserved and we have equilibrium 
between the different generations, $\m_{li} \equiv \m_l$, 
$\m_{qi} \equiv \m_q$, etc. Using also the sphaleron relation and the 
hypercharge constraint, one can express all chemical potentials, and 
therefore all asymmetries, in terms of a single chemical potential that 
may be chosen to be $\m_l$,
\begin{eqnarray}\label{exam1}
\m_e &=& {2N_f+3\over 6N_f+3}\m_l\;, \quad
\m_d = -{6N_f+1\over 6N_f+3}\m_l\;, \quad
\m_u = {2N_f-1\over 6N_f+3}\m_l\;, \nonumber\\
\m_q &=& -{1\over 3} \m_l\;, \quad
\m_{H} = {4N_f\over 6N_f+3} \m_l\;.
\end{eqnarray}
The corresponding baryon and lepton asymmetries are
\begin{equation}
B = -{4N_f\over 3}\m_l\;, \quad L = {14N_f^2+9N_f\over 6N_f+3}\m_l\;.
\end{equation}
This yields the important connection between the $B$, $B-L$ and $L$ 
asymmetries \cite{ks}
\begin{equation}\label{connection}
B = c_s (B-L); ~~L= (c_s - 1)(B- L) \;,
\end{equation}
where $c_s = (8N_f+4)/(22N_f+13)$.  The above relations hold for temperatures 
$T\gg v_F$. In general, the ratio $B/(B-L)$ is a function of $v_F/T$  
\cite{ls00}.

The relations (\ref{connection}) between B-, (B-L)- and L-number 
suggest that (B-L)-violation is needed in order to 
generate a B-asymmetry.\footnote{In the case of Dirac neutrinos,
which have extremely small Yukawa couplings, one can construct Leptogenesis 
models where an asymmetry of lepton doublets is accompanied
by an asymmetry of right-handed neutrinos such that the total L-number
is conserved and the (B-L)-asymmetry vanishes \cite{dlx}.}
Because the (B-L)-current has no anomaly, the value of B-L at time $t_f$,
where the Leptogenesis process is completed, determines the value of the
baryon asymmetry today,
\begin{equation}
B(t_0)\ =\ c_s\ (B-L)(t_f)\;.
\end{equation}
On the other hand, during the Leptogenesis process the strength of (B-L)-,
and therefore L-violating interactions can only be weak. Otherwise, because 
of Eq. (\ref{connection}), they would wash out any baryon asymmetry. 
As we shall 
see in the following, the interplay between these conflicting 
conditions leads to important constraints on the properties of neutrinos.

\section{Experimental and Theoretical Information on the Neutrino
Sector}

\subsection{Results from Oscillation Experiments}

The search for neutrino mass has  a long history \cite{history}. Positive results
are now provided by neutrino oscillation experiments. The allowed values 
for the mass-squared differences $\Delta m_{ij}^2$ and the mixing angles 
$\theta_{ij}$ at the $3\sigma$ level for three generations of neutrinos are summarized below \cite{PDG}: 
\begin{equation}
{\rm sin}^22\theta_{23}=0.92-1,~~ |\Delta
m^2_{23}|= (1.2-4.8)\times 10^{-3} {\rm eV}^2 ~~({\rm atmospheric}~ \nu)
\end{equation}
\begin{equation}
{\rm sin}^22\theta_{12}=0.70-0.95,~~ |\Delta m^2_{12}|= 
(5.4-9.5)\times 10^{-5} {\rm eV}^2 ~~({\rm solar}~\nu).
\end{equation}
The CHOOZ experiment \cite{CHOOZ} gives only an upper limit on the remaining mixing angle $\theta_{13}$:
\begin{equation} 
{\rm sin}^22\theta_{13}=0-0.17 ~~({\rm CHOOZ} ~\nu).
\end{equation} 
The LSND experiment \cite{LSND} reports neutrino oscillations from ${\bar \nu}_\mu$ 
to ${\bar \nu}_e$. The mixing angle and mass-squared difference inferred from this experiment are ${\rm
sin}^22\theta =0.003-0.03$ and $|\Delta m^2|= 0.2-2~ {\rm eV}^2$. This
parameter region is almost excluded by negative results from a
comparable experiment by the KARMEN collaboration \cite{KARMEN}, 
but there still
remains a narrow region allowed at the $90 \%$ CL.

It is very difficult to explain all the above data from neutrino oscillation
experiments within the three neutrino framework. Indeed,
to accommodate the LSND data one must introduce at least one sterile neutrino
\cite {sterile}, or make the radical assumption that CPT is not conserved 
\cite{CPT}. In this review, for simplicity we will disregard
the data from the LSND experiment. 

\subsection{Information from $\beta$-decay, 2$\beta$-decay and
Cosmology}

The oscillation experiments discussed in the previous subsection are
only sensitive to mass-squared differences. In this subsection we quote
results from direct searches for the absolute values of neutrino masses.

The direct laboratory limits on the neutrino masses are summarized as
follows \cite{PDG}:
\begin{equation}
m_{\nu_e} < 2.5 ~{\rm eV} ~;
\end{equation}
\begin{equation}
m_{\nu_\mu} < 170 ~{\rm keV} ~;
\end{equation}
\begin{equation} 
m_{\nu_\tau} < 18 ~{\rm MeV} ~.
\end{equation} 
The $\nu_e$ mass measurements use the decay of tritium, $^3{\rm H}\rightarrow
^3{\rm He} + {\bar \nu}_e + e^-$, which has a small $Q$ value, $Q=18.6$ keV,
and looks at the electron spectrum near the end point in the Kurie plot.
The limit on the $\nu_\mu$ mass is obtained from the two-body kinematics
of the pion decay, $\pi^+\rightarrow \mu^+ + \nu_\mu$. Finally, the limit on the
$\nu_\tau$ mass is obtained from measurements of the invariant mass distribution of
$3\pi$ and $5\pi$ systems in the $\tau$ decays, $\tau\rightarrow 3(5)\pi
+ \nu_\tau$.

Neutrinoless double $\beta$ decay experiments provide a bound 
on an element of the Majorana mass matrix, $m_{\nu_{ee}}$ \cite{Kayser}. The best limit
comes from the $^{76}{\rm Ge}$ results. Because the calculation of the double
$\beta$ decay rate is model dependent, we quote a range for this bound:
\begin{equation}
m_{\nu_{ee}} < 0.3-0.8 ~{\rm eV}~.
\end{equation}

One can derive a stringent upper limit on the sum of neutrino masses from 
cosmology. In the early Universe neutrinos were in thermal
equilibrium with radiation and one can infer their number density today  
to be $n_{i} \simeq 110~{\rm cm}^{-3}$ for each neutrino species. Although the contribution of massless neutrinos to the Universe's energy density is negligible, the contribution of
massive neutrinos could be important. By
requiring that the energy density of massive neutrinos does not exceed that of dark matter, one
obtains the bound:
\begin{equation}
\sum_{i}m_{i} <30h^2 ~{\rm eV},
\end{equation}
where $h$ is the Hubble constant in units of $100~{\rm km~s}^{-1}{\rm
Mpc}^{-1}$ 
[$h=0.71^{+0.04}_{-0.03}$ \cite{WMAP}].

Even if neutrinos satisfy the above limit, massive neutrinos would
affect the formation of cosmic structure, because the free streaming of
neutrinos suppresses density fluctuations at small scales.
The normalization of large- and small-scale fluctuations constrains the
contribution from neutrinos. Recent detailed analyses \cite{HR} 
lead to the bound:
\begin{equation}
\sum_{i}m_{i} <0.65 ~{\rm eV}.
\end{equation}
It is interesting that a similar constraint, $\sum_{i} m_{i} <2.0 ~{\rm
eV}$, has been obtained by using the cosmic microwave background data alone 
\cite{fukugita}.

\subsection{The Seesaw Mechanism}

If there are right-handed neutrinos, then neutrinos can have a Dirac mass much as quarks and charged leptons do. However, if this is the only source for their mass, it is not easy to find a natural
reason for the very small mass of neutrinos. With only a Dirac mass term, the smallness of 
neutrino masses needs to be ascribed to having very tiny Yukawa coupling constants $h$ ($h \sim
10^{-13}$ gives a neutrino mass $m_{\nu}\simeq 0.01$ eV). Although it is possible to imagine mechanisms that result in very small Dirac masses for 
neutrinos,\footnote{ At the end of this subsection
we outline a recent approach that may explain how such extremely small Yukawa coupling constants might arise in a higher dimensional theory.}
the seesaw mechanism, which entails Majorana masses, 
provides a natural explanation for the smallness of neutrino masses in
theories of unification at ultra-high energy scales \cite{seesaw}. 

To appreciate this point,  we should first note that the Standard Model does not require  neutrinos to be massive, because right-handed neutrinos 
are not necessary for the electroweak theory. Without right-handed neutrinos, 
these particles may acquire their mass only from what are called irrelevant 
operators -- operators such as 
$\ell\ell HH$ with dimension greater than four. These operators can give rise to a Majorana mass for neutrinos in theories with a cutoff. However, in the
limit of an infinite cutoff, neutrino masses vanish. It should be noted that if
one adopts the Planck scale as the Wilson cutoff for the 
Standard Model, one finds neutrino masses to be at most $10^{-5}$ eV. 
Thus, the observed mass $\sqrt{\Delta m^2_{\rm atm}}\simeq 0.05$ eV is 
unable to be explained within the Standard Model.

If one considers possible extensions of the Standard Model gauge group, 
it is natural to consider a gauge group $G$ whose rank is at least 5, 
since the rank of the Standard Model group,
$SU(3)\times SU(2)\times U(1)_{Y}$, is 4. This new group $G$ may
contain an extra $U(1)$ as a subgroup, in addition to the Standard Model 
gauge group. The simplest candidate for the 
extra $U(1)$ is a B-L gauge symmetry, because we know that the global
$U(1)_{B-L}$ does not have any anomaly due to the Standard Model gauge
interactions.  However, when one gauges this B-L symmetry,
 the theory does have
a self-anomaly of the B-L\\ interactions. That is, the triangle anomaly
of $[U(1)_{B-L}]^3$ is non-vanishing. A crucial point, however, is that this
anomaly is cancelled by introducing right-handed neutrinos. This also
cancels the mixed gravitational/B-L anomaly. Thus,
right-handed neutrinos are required for consistency of the theory! A famous
example which includes $U(1)_{B-L}$ is provided by $SO(10)$ grand unification.
But our argument is more general. For instance, the string brane world
predicts many $U(1)$'s and it is quite natural to consider some of them
to be anomaly free and survive as low-energy (compared with the string
scale) gauge symmetries.

It is usually assumed that the unification group $G$ is broken
down to the Standard Model group at high energies. Then, the right-handed
neutrinos naturally obtain large Majorana masses, because they are singlets 
of the Standard Model and there is no unbroken symmetry to 
protect them from acquiring a large Majorana mass. In this article we shall denote  heavy
Majorana neutrinos as $N$.
Then, the masses of the neutrinos written as a matrix take a simple form:
\begin{equation}
\left(
\begin{array}{cc}
0&m \\
m^T&M
\end{array}
\right)\;.
\end{equation}
Here $m$ is the Dirac mass matrix between the left-handed neutrinos $\nu$ and 
the heavy Majorana neutrinos $N$,
which is of order the electroweak scale, while $M$ is the Majorana mass matrix of the heavy neutrinos $N$. For three generations, both $M$ and $m$ are $3 \times 3$ matrices.
Integrating out the heavy Majorana neutrinos $N$ leads to a small neutrino
mass via the seesaw mechanism \cite{seesaw}. For one neutrino generation one simply has that: 
\begin{equation}
m_\nu \simeq \frac{m^2}{M}.
\end{equation}
We see from the above that a small neutrino mass is a reflection of the ultra-heavy mass
of the heavy neutrino $N$. The observed neutrino mass
$\sqrt{\Delta m^2_{\rm atm}}\simeq 0.05$ eV implies $M\simeq 10^{15}$ GeV,
which is very close to the Grand Unification scale. Thus, effectively, the small neutrino masses provide a window to  new physics at
an ultra-high energy scale.

One may question why the unification group $G$, or the $U(1)_{B-L}$ symmetry, 
should be broken at such a high energy scale. Perhaps the answer to this
question, as we shall amplify in this review, is because, otherwise, the baryon number in the Universe would be too small
for us to exist! It turns out that if the
Universe's baryon number were to be two orders of
magnitude below the present observed value galaxies would not be formed 
\cite{TR97}. One of the purposes 
of this review article is to explain in some detail this fundamental 
point.\\

\noindent
{\bf 3.3.1 Small Neutrino Yukawa Couplings from Higher Dimensional Theories}\\
\noindent
To explain how one can generate small Dirac masses for neutrinos,  consider a theory described in (4+1)-dimensional space-time,
while our world is on a (3+1)-dimensional hyperplane -- a so-called D3 brane.
The Einstein action of gravity in five-dimensional space-time is given
by
\begin{equation}
S= \frac{M_*^3}{16\pi}\int d^4x\int dy\sqrt{-g_5}{\cal R}_5,
\end{equation}
where $M_*$ is the gravitational scale in five-dimensional space-time,
and $g_5$ and ${\cal R}_5$ are the metric and the scalar curvature,
respectively. We assume that the fifth dimension is compactified to a
space of radius $L$, and consider the metric to be
\begin{equation}
d^2s = g_{\mu\nu}dx^\mu dx^\nu -dy^2,
\end{equation}
where $g_{\mu\nu}$ is the metric in four-dimensional space-time. The
integration over $dy$ leads to the four-dimensional action
\begin{equation}
S_4 = \frac{M_*^3L}{16\pi}\int d^4x\sqrt{-g_4}{\cal R}_4.
\end{equation}
Then, the Planck scale, $M_{\rm P}\simeq 1.2\times 10^{19}$ GeV, in 
four-dimensional space-time is given by
\begin{equation}
M_{\rm P}^2 =  M_*^3L.
\end{equation}
One can get the observed value for $M_{\rm P}$ even for $M_* = 1$ TeV
by taking a very large $L$.
The weakness of gravity in these theories is the result of having a large compactification scale in
the fifth dimension \cite{arkani}.

Let us now assume that all the Standard Model particles reside on a D3 brane
at the boundary $y=0$ and that the right-handed neutrino lives in the
five-dimensional bulk \cite{arkani2}. Then the action involving the
right-handed neutrinos is given by
\begin{equation}
S= M_*\int d^4x\int^L_0 dy\sqrt{-g_5}{\bar N_R} i \partial\llap{/} N_R
+ \int d^4x\int^L_0 dy\sqrt{-g_5}h{\bar N_R}\ell_LH\delta (y) + 
{\rm h.c.}\,, 
\end{equation}
Here, $\ell_L$ and $H$ are SU(2)$_L$ doublets of the left-handed leptons
and of the Higgs boson, respectively. One obtains the action in four
dimensions by integrating over $dy$, and one finds:
\begin{equation}
S_4 = M_*L\int d^4x\sqrt{-g_4}{\bar N_R} i \partial\llap{/} N_R
+ \int d^4x\sqrt{-g_4}h{\bar N_R}\ell_LH + {\rm h.c.}\,. 
\end{equation}
After renormalizing the wave function of $N_R$ so that it has a
canonical kinetic term, one finds that the Yukawa coupling constant is
suppressed by $1/\sqrt{M_*L} = M_*/M_{\rm P}$. Thus in this model one obtains  an effective
Yukawa coupling constant $h_{\rm eff} \simeq 10^{-13}$ (corresponding to a neutrino
Dirac mass  $m_{\nu}\simeq 0.01$ eV ) for $h=1$ and $M_* \simeq 10^3$ TeV.

This model, however, has a serious drawback. There is no symmetry to
protect the right-handed neutrinos from acquiring a Majorana mass, because
they  are singlets of the Standard Model gauge group. A solution to this
problem may be found by imposing a $U(1)_{B-L}$ gauge symmetry in
the five-dimensional bulk. As long as the B-L gauge symmetry is exact,
the right-handed neutrinos cannot have a Majorana mass because
this mass carries a non-vanishing B-L charge. If this symmetry is exact, the corresponding gauge boson is completely
massless. However, this may not cause any phenomenological difficulties at 
low-energies, because the B-L gauge coupling constant must also be extremely
suppressed.\footnote{The B-L gauge coupling constant $\alpha_{B-L}$ 
is constrained to be $\alpha_{B-L} < 10^{-21}\alpha_{\rm em}$. This bound 
comes from the empirical limits of the electromagnetic charges for the
neutron and the neutrino. That is, $Q_n = (-0.4 \pm 1.1)\times 10^{-21}$ 
\cite{chargenu} and $Q_\nu = (0.5\pm2.9)\times 10^{-21}$ \cite{chargeneu}.
The present model suggests $\alpha_{B-L}\simeq (\frac{M_*}{M_P})^2\times\alpha_{\rm
em}\simeq 10^{-25}\times\alpha_{\rm
em}$, which may be in an interesting region for future
experiments.} But, as we explained in Section 2, if the B-L gauge symmetry is 
exact, it is very
difficult to account for the baryon-number asymmetry in the present
Universe.

\subsection{CP-Violating Phases at Low and High Energies in the \\ Lepton
Sector}

In the Standard Model, the Lagrangian for the lepton sector, augmented by including right handed neutrinos, is given by
\begin{eqnarray} \label{llag}
{\cal L} &=& {\bar \ell}_{Li} i \partial\llap{/}\ell_{Li} 
+ {\bar e}_{Ri} i \partial\llap{/}e_{Ri}
+ {\bar N}_{Ri} i \partial\llap{/}N_{Ri} \nonumber\\
&& + f_{ij}{\bar e}_{Ri}\ell_{Lj}H^{\dagger}
+h_{ij}{\bar N}_{Ri}\ell_{Lj}H - \frac{1}{2}M_{ij}N_{Ri}N_{Rj} + {\rm h.c.}\;, 
\end{eqnarray}
where $i,j = \{1-3\}$ are the family-number indices. We adopt, without losing
generality, a basis where the matrices $f_{ij}$ and $M_{ij}$ are diagonal.
The Yukawa matrix $h_{ij}$ in this basis is in general complex and thus has CP-violating phases. Because for three families,  the matrix
$h_{ij}$ has 9 complex parameters, we have 9 possible CP-violating phases. However, three of
these phases can be absorbed into the wave function of $\ell_L$ and hence 6
CP-violating  phases remain physically relevant. These are known as
high-energy phases, because they enter in the full theory.\footnote{In 
particular, as we will show in the next Section, the phase contributing to the generation
of the lepton asymmetry in the decay of $N_1$ is a combination of these
high-energy phases, given by $\sum{\rm Im}[(hh^{\dagger})_{1i}]^2$.}

Let us now discuss the CP-violating phases at low energies. To do that we first
need to integrate out the heavy Majorana neutrinos, $N_i$. Doing so the effective
Lagrangian for the lepton sector reduces to:
\begin{equation}
{\cal L}_{\rm eff} = {\bar \ell}_{Li} i \partial\llap{/}\ell_{Li} + 
{\bar e}_{Ri} i \partial\llap{/}e_{Ri} 
+ f_{ii}{\bar e}_{Ri}\ell_{Li}H^{\dagger}
+ \frac{1}{2}\sum_{k}h^T_{ik}h_{kj}\ell_{Li}\ell_{Lj}\frac{H^2}{M_k} + 
{\rm h.c.}\,. 
\end{equation}
The last term can be rewritten as 
\begin{equation}
- \frac{1}{2}m_{\nu_{ij}}\ell_{Li}\ell_{Lj}\frac{H^2}{\langle H\rangle^2}\,,
\end{equation}
so that all low-energy phases appear in the mass matrix of light neutrinos.
Because the neutrino mass matrix is symmetric, it has 6 complex parameters, and hence one has 6 possible CP-violating phases. However,  as before,
3 of these 6 phases can be absorbed into the wave function of
$\ell_L$. Therefore, there remain only three physical low energy CP-violating 
phases \cite{petcov}. Because they are different in number,
it is unfortunately very difficult to establish a
direct link between the low-energy and the high-energy CP-violating 
phases \cite{BR}.

Furthermore, in practice, it is not possible to measure all three low-energy phases.
One of these three phases can be measured by neutrino oscillation experiments, while neutrinoless double $\beta$ decay, if it were to be observed, would provide information on another
phase. However, the remaining phase is undetermined. In other
words, one cannot perform a complete experiment to determine the
neutrino mass matrix. Nevertheless, if the neutrino mass matrix were to have an extra
constraint, one may be able to determine all matrix elements of $m_\nu$. This 
constraint must be independent of the frame of the family basis. 
One example of such a constraint is the requirement that ${\rm det}(m_{\nu})=0$. In this case, we have only 
7 independent physical parameters including the phases in the neutrino 
mass matrix \cite{branco}, which can be determined in principle in future 
experiments.\footnote{As will be shown in Section 6, Affleck-Dine Leptogenesis
suggests a constraint, $m_{\nu_1}\simeq 10^{-9}$ eV and hence ${\rm
det}(m_\nu)\simeq 0$.}

\section{Thermal Leptogenesis}

\subsection{Lepton Number Violation and Leptogenesis}

As we discussed above, lepton number violation is most simply realized by adding right-handed 
neutrinos to the Standard Model. Their existence is predicted by all 
extensions of the Standard Model containing B-L as a local symmetry
and allows for an elegant explanation of the smallness of the light neutrino 
masses via the seesaw mechanism \cite{seesaw}.

The most general Lagrangian for couplings and masses of charged leptons and 
neutrinos is given in Eq. (\ref{llag}).
The vacuum expectation value of the Higgs field, $\VEV{H}=v_F$, 
generates Dirac masses $m_e$ and $m_D$ for charged leptons and neutrinos, 
$m_e=f v_F$ and $m_D=hv_F$, which are assumed to be much
smaller than the Majorana masses $M$. This yields the light and heavy neutrino
mass eigenstates
\begin{equation}
     \n\simeq V_{\nu}^T\n_L+\n_L^c V_{\nu}^*\quad,\qquad
     N\simeq N_R+N_R^c\, ,
\end{equation}
with masses
\begin{equation}
     m_{\n}\simeq- V_{\nu}^Tm_D^T{1\over M}m_D V_{\nu}\,
     \quad,\quad  m_N\simeq M\, .
     \label{seesaw}
\end{equation}
In a basis where the charged lepton mass matrix $m_e$ and the Majorana
mass matrix $M$ are diagonal,
$V_{\nu}$ is the mixing matrix in the leptonic charged current.

The right-handed neutrinos can efficiently erase any pre-existing lepton asymmetry
at temperatures $T > M$, but they can also generate a lepton asymmetry 
by means of their out-of-equilibrium decays at temperatures $T < M$. 
This asymmetry is then partially transformed into a baryon asymmetry by sphaleron processes. This is the Leptogenesis mechanism proposed by 
Fukugita and Yanagida \cite{FY}.

The decay width of the heavy neutrino $N_i$  at tree level reads,
\begin{equation}
\G_{Di}=\G\left(N_i\to H + \ell_L\right)
+\G\left(N_i\to H^{\dagger} + \ell_L^{\dagger}\right)
           ={1\over8\p}(h h^\dg)_{ii} M_i\;.
\label{decay}
\end{equation}
Once the temperature of the universe drops below the mass $M_1$, the heavy neutrinos are not able to follow the rapid change of the equilibrium 
distribution. Hence, the necessary deviation from thermal equilibrium ensues as a result of having a too large number density of heavy neutrinos, compared to the
equilibrium density. 
Eventually, however, the heavy neutrinos decay, and a lepton asymmetry is 
generated owing to the presence of CP-violating processes. The CP asymmetry involves the interference 
between the tree-level amplitude and the one-loop vertex and self-energy
contributions (see Fig.~(1)). 
In a basis, where the right-handed neutrino mass matrix $M$
is diagonal, one obtains \cite{l-asymmetry} for the CP asymmetry parameter
$\ve_1$ assuming hierarchical heavy neutrino masses ($M_1 \ll M_2, M_3$):
\begin{equation}\label{eps}
    \ve_1 \simeq {3\over16\pi}\;{1\over\left(h h^\dg\right)_{11}}
    \sum_{i=2,3}\mbox{Im}\left[\left(h h^\dg\right)_{i1}^2\right]
    {M_1\over M_i}\; .
  \end{equation}
In the case of mass differences of order the decay widths, one obtains a 
significant enhancement from the self-energy contribution \cite{pil99},
although the influence of the thermal bath on this effect is presently unclear.

The CP asymmetry of Eq. (\ref{eps}) can be obtained in a very simple way by first 
integrating out the heavier neutrinos $N_2$ and $N_3$ in the leptonic Lagrangian. This yields
\begin{equation}\label{yuk2}
\cl_{\n}^{eff} = h_{ 1j}\Bar{N_R}_1 \ell_{Lj} H
          -{1\over2}M_1 \Bar{N^c_R}_1 N_{R1} 
          +{1\over 2} \eta_{ij} \ell_{Li} H \ell_{Lj} H + \mbox{ h.c.}\;,
\end{equation}
with
\begin{equation}
\eta_{ij} = \sum_{k=2}^3 h^T_{ ik}{1\over M_k} h_{ kj}\;. 
\end{equation}
The asymmetry $\ve_1$ is then obtained from the interference of the
Born graph and the one-loop graph involving the cubic and the quartic 
couplings. 
 \begin{figure}[t]
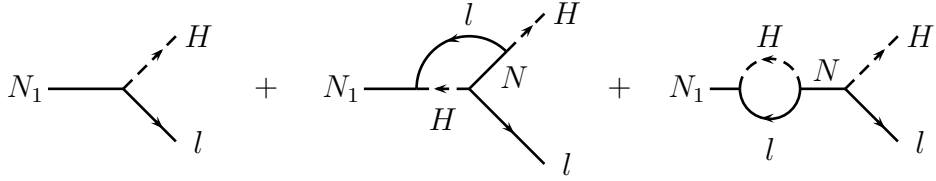

    \centerline{\parbox[c]{12.5cm}{
\pspicture(0,0)(3.7,2.6)
\psline[linewidth=1pt](0.6,1.3)(1.6,1.3)
\psline[linewidth=1pt](1.6,1.3)(2.3,0.6)
\psline[linewidth=1pt,linestyle=dashed](1.6,1.3)(2.3,2.0)
\psline[linewidth=1pt]{->}(2.03,0.87)(2.13,0.77)
\psline[linewidth=1pt]{->}(2.03,1.73)(2.13,1.83)
\rput[cc]{0}(0.3,1.3){$N_1$}
\rput[cc]{0}(2.6,0.6){$l$}
\rput[cc]{0}(2.6,2.0){$H$}
\rput[cc]{0}(3.5,1.3){$+$}
\endpspicture
\pspicture(-0.5,0)(4.2,2.6)
\psline[linewidth=1pt](0.6,1.3)(1.3,1.3)
\psline[linewidth=1pt,linestyle=dashed](1.3,1.3)(2.0,1.3)
\psline[linewidth=1pt](2,1.3)(2.5,1.8)
\psline[linewidth=1pt,linestyle=dashed](2.5,1.8)(3,2.3)
\psline[linewidth=1pt](2,1.3)(3,0.3)
\psarc[linewidth=1pt](2,1.3){0.7}{45}{180}
\psline[linewidth=1pt]{<-}(1.53,1.3)(1.63,1.3)
\psline[linewidth=1pt]{<-}(1.7,1.93)(1.8,1.96)
\psline[linewidth=1pt]{->}(2.75,2.05)(2.85,2.15)
\psline[linewidth=1pt]{->}(2.5,0.8)(2.6,0.7)
\rput[cc]{0}(0.3,1.3){$N_1$}
\rput[cc]{0}(1.65,0.9){$H$}
\rput[cc]{0}(2,2.3){$l$}
\rput[cc]{0}(2.6,1.45){$N$}
\rput[cc]{0}(3.3,2.3){$H$}
\rput[cc]{0}(3.3,0.3){$l$}
\rput[cc]{0}(4.0,1.3){$+$}
\endpspicture
\pspicture(-0.5,0)(3.5,2.6)
\psline[linewidth=1pt](0.5,1.3)(0.9,1.3)
\psline[linewidth=1pt](1.7,1.3)(2.3,1.3)
\psarc[linewidth=1pt](1.3,1.3){0.4}{-180}{0}
\psarc[linewidth=1pt,linestyle=dashed](1.3,1.3){0.4}{0}{180}
\psline[linewidth=1pt]{<-}(1.18,1.69)(1.28,1.69)
\psline[linewidth=1pt]{<-}(1.18,0.91)(1.28,0.91)
\psline[linewidth=1pt](2.3,1.3)(3.0,0.6)
\psline[linewidth=1pt,linestyle=dashed](2.3,1.3)(3.0,2.0)
\psline[linewidth=1pt]{->}(2.73,0.87)(2.83,0.77)
\psline[linewidth=1pt]{->}(2.73,1.73)(2.83,1.83)
\rput[cc]{0}(1.3,0.5){$l$}
\rput[cc]{0}(1.3,2){$H$}
\rput[cc]{0}(2.05,1.55){$N$}
\rput[cc]{0}(0.2,1.3){$N_1$}
\rput[cc]{0}(3.3,0.6){$l$}
\rput[cc]{0}(3.3,2.0){$H$}
\endpspicture
}}
\caption{Tree level and one-loop diagrams contributing to heavy neutrino
decays whose interference leads to Leptogenesis.}
  \end{figure}
%
%
This includes automatically both, vertex and self-energy
corrections \cite{bf00} and yields an expression for $\ve_1$ directly
in terms of the light neutrino mass matrix:
\begin{equation}\label{CPas}
\ve_1 \simeq -{3\over 16\p}{M_1\over (h h^\dg)_{11} v_F^2} 
{\rm Im}\left(h^* m_\n h^\dg\right)_{11}\;. 
\end{equation}

The CP asymmetry then leads to a  (B-L)-asymmetry \cite{FY},
 \begin{equation}\label{basym}
    Y_{B-L}\simeq -Y_L = - \frac{n_L-n_{\Bar{L}}}{ s}\ = -\k\frac{\ve_1}{g_*}\;.
  \end{equation}
Here $s$ is the entropy and, in the present epoch $s=7.04 n_\gamma$, whereas $g_*\sim 100$ is the number of degrees of freedom in the plasma.
The factor $\k<1$ in the above takes into account the effect of washout processes. As we shall discuss below, in order to 
determine $\k$ one has to solve the Boltzmann equations.

Early studies of Leptogenesis were partly motivated by trying to find alternatives to Electroweak Baryogenesis, which did not seem to produce a big enough asymmetry. Some extensions of the Standard Model were considered and, in particular, in the 
simple case of hierarchical heavy neutrino masses the observed value of the 
baryon asymmetry is naturally obtained with B-L broken at the unification 
scale, $M_{GUT} \sim 10^{15}$~GeV. The corresponding light neutrino masses
are then very small, $m_{1,2} < m_{3} \sim 0.1$~eV, and the typical 
parameters for the necessary 
CP asymmetry and the Baryogenesis temperature are $\ve_1 \simeq 10^{-6}$ and
$T_B \sim M_1 \sim 10^{10}\ {\rm GeV}$, respectively 
\cite{bp96,by99}.\footnote{For early work based on $SO(10)$, see \cite{Ghe}.}
 Subsequently, researchers realized that
such small neutrino masses are consistent with the small mass differences
inferred from the solar and atmospheric neutrino oscillations.
This fact has given rise to a strong interest in Leptogenesis in recent years, and
a large number of interesting models have been suggested \cite{models}.

\subsection{Departure from Thermal Equilibrium}

Leptogenesis takes place at temperatures $T \sim M_1$. For a decay width small
 compared to the Hubble parameter, $\G_1(T) < H(T)$, heavy neutrinos are out 
of thermal equilibrium, otherwise they are in thermal equilibrium \cite{KT}. 
The 
borderline between the two regimes is given by $\G_1 = H|_{T=M_1}$, which
is equivalent to the condition that the effective neutrino mass 
\beqa\label{effm}
\mt = {(m_D m_D^\dg)_{11}\over M_1} 
\eeqa
is equal to the `equilibrium neutrino mass'
\beq\label{mequ}
m_* = {16\p^{5/2}\over 3\sqrt{5}} g_*^{1/2} {v_F^2\over M_{\rm P}} 
\simeq 10^{-3}~\mbox{eV}\;.
\eeq
Here we have used the Hubble parameter $H(T)\simeq 1.66\,g_*\,T^2/M_{\rm P}$
where $g_*=g_{SM}=106.75$ is the total number of degrees of freedom and
$M_{\rm P}=1.22\times10^{19}\,{\rm GeV}$ is the Planck mass. 

It is quite remarkable that the equilibrium neutrino mass $m_*$ is close to the
neutrino masses suggested by neutrino oscillations,
$\sqrt{\D m^2_{\rm sol}} \simeq 8\times 10^{-3}$~eV and 
$\sqrt{\D m^2_{\rm atm}} \simeq 5\times 10^{-2}$~eV.
This encourages one to think that it may be possible to understand the cosmological baryon 
asymmetry via Leptogenesis as a process close to thermal equilibrium. Ideally,
$\D L=1$ and $\D L=2$ processes should be strong enough at temperatures above 
$M_1$ to keep the heavy neutrinos in thermal equilibrium and weak enough to 
allow the generation of an asymmetry at temperatures below $M_1$.  

In general, the generated baryon asymmetry is the result of a competition 
between production processes and washout processes that tend to erase any 
generated asymmetry. Unless the heavy Majorana neutrinos are partially 
degenerate, $M_{2,3}-M_1 \ll M_1$, the dominant processes are decays and 
inverse decays of $N_1$ and the usual off-shell $\D L=1$ and $\D L=2$ 
scatterings \cite{lut,plu}. 

\begin{figure}[t]        
\centerline{\psfig{file=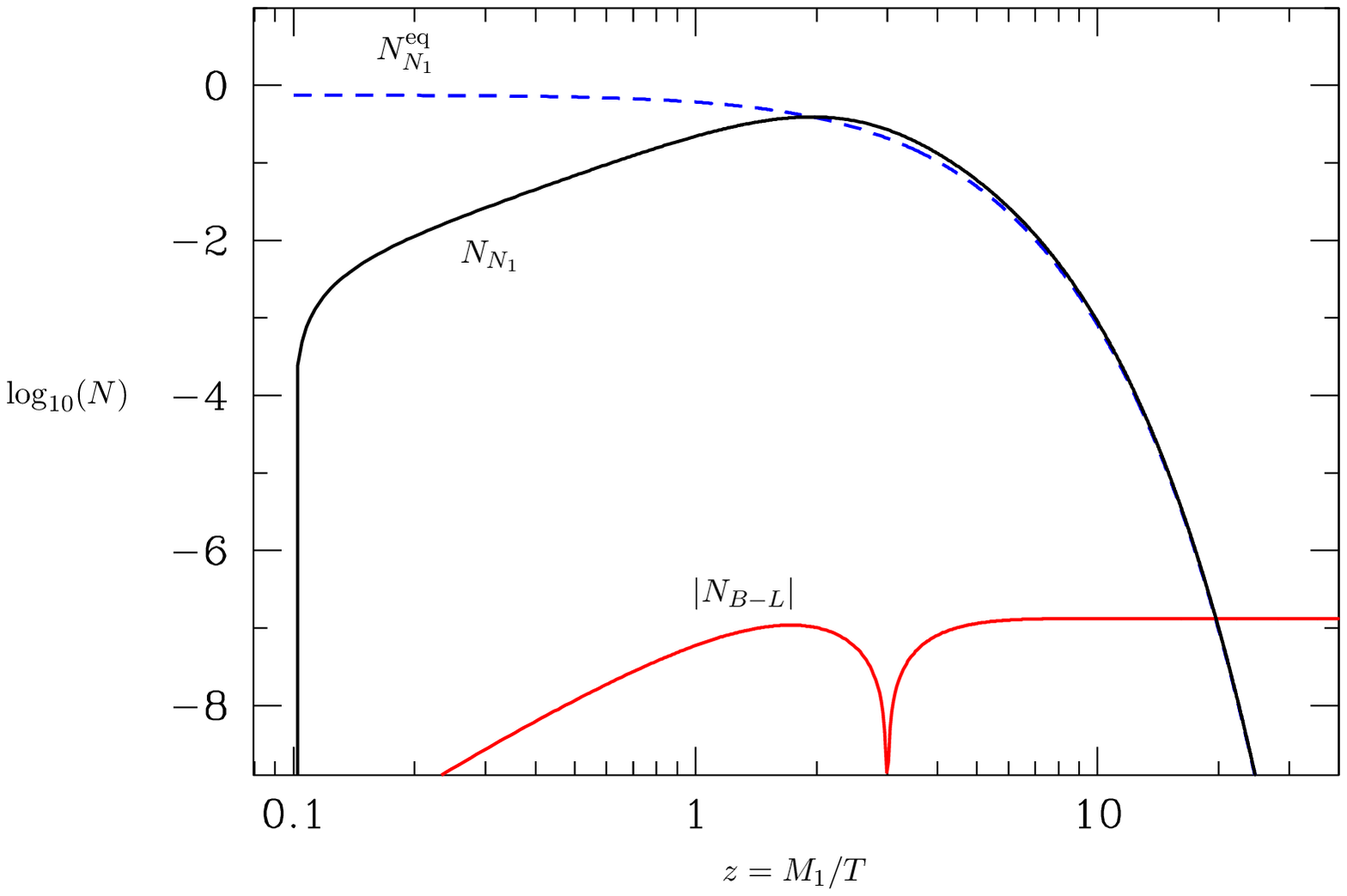,height=6.5cm,width=11cm}}
\caption{The evolution of the $N_1$ abundance and the $B-L$ asymmetry
for a typical choice of parameters, $M_1=10^{10}\,$GeV, 
$\varepsilon_1=10^{-6}$, $\widetilde{m}_1=10^{-3}\,$eV
and $\overline{m}=0.05\,$eV. From \cite{bdp02}.}
\label{BASYM}
\end{figure}

The Boltzmann equations for Leptogenesis are\footnote{We use the
conventions of \cite{bdp02}. We have also summed over the three lepton 
flavours neglecting the dependence on the lepton Yukawa couplings 
\cite{bcx00}.}
\beqa
{dN_{N_1}\over dz} & = & -(D+S)\,(N_{N_1}-N_{N_1}^{\rm eq}) \;, \label{lg1} \\ 
{dN_{B-L}\over dz} & = & -\ve_1\,D\,(N_{N_1}-N_{N_1}^{\rm eq})-W\,N_{B-L} \;,
\label{lg2}
\eeqa
where $z=M_1/T$. The number density $N_{N_1}$ and the amount of $B-L$ 
asymmetry, $N_{B-L}$, are calculated in a portion of comoving volume that 
contains one photon at the onset of Leptogenesis, so that the relativistic 
equilibrium $N_1$ number density is given by $N_{N_1}^{\rm eq}(z\ll 1)=3/4$. Alternatively, one may normalize the number density to the entropy density
$s$ and consider $Y_X = n_X/s$. If entropy is conserved, both normalizations are related by a constant. 

There are four classes of processes that contribute to the different terms in
the above equations: decays, inverse decays, $\D L=1$ scatterings and $\D L=2$ 
processes mediated by heavy neutrinos. The first three processes all modify the
$N_1$ abundance and try to push it towards its equilibrium value 
$N_{N_1}^{\rm eq}$. Denoting by $H$ the Hubble expansion rate, the term 
$D = \Gamma_D/(H\,z)$ accounts for decays and inverse decays, whereas the 
scattering term $S = \Gamma_S/(H\,z)$ represents the $\D L=1$ scatterings. 
Decays also yield the source term for the generation of the $B-L$ asymmetry, 
the first term in Eq.~(\ref{lg2}), whereas all other processes
contribute to the total washout term $W = \Gamma_W/(H\,z)$ which competes
with the decay source term. The dynamical generation of the $N_1$ abundance
and the $B-L$ asymmetry is shown in Fig. (\ref{BASYM}) for typical parameters.

\subsection{Decays and Inverse Decays}

It is very instructive to consider first a simplified picture in which decays 
and inverse decays are the only processes that are effective.\footnote{This section follows 
closely \cite{bdp04}.} For consistency, in this approximation the real 
intermediate state contribution to the $2\rightarrow 2$ processes has to be 
included. In the kinetic equations (\ref{lg1}) and (\ref{lg2}) one then has 
to replace $D+S$ by $D$ and $W$ by $W_{I\!D}$, respectively, where $W_{I\!D}$ 
is the contribution of inverse decays to the washout term.
The solution for $N_{B-L}$ in this case is the sum of two terms \cite{KT},
\begin{equation}\label{solution}
N_{B-L}(z)=N_{B-L}^{\rm i}\,e^{-\int_{z_{\rm i}}^{z}\,dz'\,W_{I\!D}(z')}
-{3\over 4}\,\ve_1\,\k(z;\mt)\;.
\end{equation}
Here the first term accounts for an initial asymmetry which is partly reduced 
by washout, and the second term describes $B-L$ production from $N_1$ decays.
It is expressed in terms of the {\em efficiency factor} $\k$  \cite{bcx00} 
which does not depend on the CP asymmetry $\ve_1$, 
\begin{equation}\label{lg3}
\k(z)={4\over 3} \int_{z_{\rm i}}^{z}\,dz'\,
D \left(N_{N_1}-N_{N_1}^{\rm eq}\right)\,
e^{-\int_{z'}^{z}\,dz''\,W_{I\!D}(z'')}\;.
\end{equation}
As we shall see, decays and inverse  decays are sufficient to describe qualitatively many properties of the full problem.

We will first study in 
detail the regimes of weak and strong washout. If just decays and inverse 
decays are taken into account, these regimes correspond, respectively,  to the limits $K\ll 1$ and $K\gg 1$ of the decay parameter
\begin{equation}\label{decpar}
K = {\G_D(z=\infty)\over H(z=1)} = {\widetilde{m}_1\over m_*} \;,
\end{equation}
introduced in the context of ordinary GUT baryogenesis \cite{KT}.
Based on the insight into the dynamics of the non-equilibrium process gained 
from these limiting cases one can then obtain analytic interpolation formulas 
that describe rather accurately the entire parameter range. 

To proceed, let us first recall some basic definitions and formulas.
The decay rate is given by the formula \cite{kw80},
\begin{equation}
\Gamma_D(z) = \Gamma_{D1}\,\left\langle {1\over \gamma} \right\rangle \;,
\end{equation}
where the thermally averaged dilation factor is given by the ratio of
the modified Bessel functions $K_1$ and $K_2$,
\begin{equation}\label{G}
\left\langle {1\over \gamma} \right\rangle =
{K_1(z)\over K_2(z)}\;.
\end{equation}
For the decay term $D$, one then obtains
\begin{equation}\label{D}
D(z) = K\,z\,\left\langle {1\over \gamma} \right\rangle\;.
\end{equation}

The inverse decay rate is related to the decay rate by
\begin{equation}
\Gamma_{I\!D}(z) =\Gamma_D(z)\,{N_{N_1}^{\rm eq}(z)\over N_l^{\rm eq}} \; ,
\end{equation}
where $N_l^{\rm eq}$ is the equilibrium density of lepton doublets. Because
the number of degrees of freedom for heavy Majorana neutrinos and lepton 
doublets is the same, $g_{N_1}=g_l=2$, one has
\begin{equation}\label{Neq}
N_{N_1}^{\rm eq}(z) = {3\over 8} z^2 K_2(z) \; , \quad
N_l^{\rm eq} = {3\over 4}\;.
\end{equation}
This yields for the contribution of inverse decays to the washout term 
$W$:
\begin{equation}\label{WID2}
W_{I\!D}(z) = {1\over 2} D(z)\,{N_{N_1}^{\rm eq}(z)\over N_l^{\rm eq}}\;.
\end{equation}
All relevant quantities are given in terms of the Bessel functions 
$K_1$ and $K_2$, which can be approximated by simple analytical expressions.

In the regime {\it far out of equilibrium}, $K \ll 1$, decays occur at 
very small temperatures, $z\gg 1$, and the produced ($B-L$)-asymmetry 
is not reduced by washout effects. In this case, using Eq. (\ref{lg1}) with $S=0$, the integral for the 
efficiency factor given in Eq. (\ref{lg3}) becomes simply,
\begin{equation}\label{oodec}
\kappa(z)\simeq{4\over 3}\left(N_{N_1}^{\rm i} - N_{N_1}(z)\right) \; .
\end{equation}
The final value of the efficiency factor $\k_{\rm f} = \k(\infty)$ is 
proportional to the initial $N_1$ abundance. If 
$N_{1}^{\rm i}=N_{1}^{\rm eq}=3/4$, then $\k_{\rm f}=1$.
But if the initial abundance is zero, then $\k_{\rm f}=0$ as well. Therefore, 
in this region there is the well known problem that one has to invoke
some external mechanism to produce the initial abundance of neutrinos.
Moreover, an initial (B-L)-asymmetry is not washed out. Thus in the regime $K \ll 1$ the results strongly depend on
the initial conditions and there is little predictivity.

In order to obtain the efficiency factor in the case of {\it vanishing 
initial $N_1$-abundance}, 
$N_{N_1}(z_{\rm i}) \equiv N_{N_1}^{\rm i} \simeq 0$, one has to
calculate how heavy neutrinos are dynamically produced by inverse decays.
This requires solving the kinetic equation Eq. (\ref{lg1}) with the initial 
condition $N_{N_1}^{\rm i} = 0$.

Let us define a value $z_{\rm eq}$ by the condition
\begin{equation}\label{condeq}
N_{N_1}(z_{\rm eq})=N_{N_1}^{\rm eq}(z_{\rm eq}) \;.
\end{equation}
Then Eq.~(\ref{lg1}) implies that the number density reaches its maximum at 
$z=z_{\rm eq}$. For $z > z_{\rm eq}$ the efficiency factor is 
always the sum of two contributions,
\begin{equation}
\k_{\rm f}(z) = \k^-(z) + \k^+(z) \;.
\end{equation}
Here $\k^-(z)$ and $\k^+(z)$ correspond to the integration domains
$[z_{\rm i},z_{\rm eq})$ and $[z_{\rm eq},z)$, respectively.

Consider first the case of {\it weak washout}, $K \ll 1$, which implies 
$z_{\rm eq} \gg 1$. One then finds,
\begin{equation}\label{nkweak}
N_{N_1}(z_{\rm eq})\simeq {9\pi \over 16}\ K \; .
\end{equation}
It turns out that to first order in $K$, there is a cancellation between
$\k^+$ and $\k^-$, yielding for the final efficiency factor
\begin{equation}\label{reduction}
\k_{\rm f}(K) 
\simeq {9\pi^2\over 64}\,K^2 \; .
\end{equation}
Note, that Eq.~(\ref{reduction}) does not hold for $K > 1$, because 
in this case
$z_{\rm eq}$ becomes small, and washout effects change the result.

In the case of {\it strong washout}, $K\gg 1$, we can neglect the negative
contribution $\k^-$, because the asymmetry generated at high temperatures
is efficiently washed out. Now the neutrino abundance tracks closely the 
equilibrium behavior. Because $D \propto K$, one can solve Eq.~(\ref{lg1}) systematically in powers of $1/K$, which yields
\begin{equation}\label{del}
D\left(N_{N_1}(z)-N_{N_1}^{\rm eq}(z)\right) = 
{3\over 2Kz} W_{I\!D}(z)+ {\cal O}\left({1\over K}\right)\; ,
\end{equation}
where we have used properties of the Bessel functions. From Eqs. 
(\ref{lg3}) and (\ref{del}) one obtains for the efficiency 
factor\footnote{Because $\k^-$ does not contribute we can take the lower limit below as $z_{\rm i}$.}
\begin{equation}\label{efd}
\k(z) = {2\over K} \int_{z_{\rm eq}}^{z} dz'\, {1\over z'} W_{I\!D}(z')\,
e^{- \int_{z'}^{z} dz''\, W_{I\!D}(z'')}\;.
\end{equation}
The integral is dominated by the contribution from a region around the value 
$z_{\rm B}$ where the integrand has a maximum, which is determined by the
condition
\begin{equation}\label{eqz0}
W_{I\!D}(z_{\rm B}) = 
\left\langle {1\over\gamma}\right\rangle^{-1}(z_{\rm B})\,
-\, {3\over z_{\rm B}}\;.
\end{equation}
For $K\gg 1$ one has $z_{\rm B}\gg 1$, and the condition (\ref{eqz0}) becomes
approximately $W_{I\!D}(z_{\rm B})\simeq 1$, with $W_{I\!D}(z) > 1$ for
$z < z_{\rm B}$ and $W_{I\!D}(z) < 1$ for $z > z_{\rm B}$. This means that
the asymmetry produced for $z < z_{\rm B}$ is essentially erased, 
whereas for $z > z_{\rm B}$, washout is negligible.
Hence, the expression of Eq. (\ref{efd}) is a good approximation for
the final efficiency factor.

One finds that a rather accurate expression for $z_B(K)$ is given by 
\begin{equation}
  z_B(K) \simeq 1+{1\over2}\,\mbox{ln}\left(1+{\pi K^2\over1024}
     \left[\mbox{ln}\left({3125\pi K^2\over1024}\right)\right]^5\right)\;.
  \label{interpol}
\end{equation}
The integral  of Eq. (\ref{efd}), which gives the final efficiency factor in
terms of $z_B(K)$, is well approximated by 
\begin{equation}\label{kfan}
\k_{\rm f}(K) \simeq
{2\over z_{\rm B}(K)K}\left(1-e^{-{1\over 2}z_{\rm B}(K)K}\right)\;.
\end{equation}
Both equations can also be extrapolated into the regime of weak washout, 
$K \ll 1$, where one obtains $\k_{\rm f} = 1$ corresponding to thermal initial
abundance, $N_{N_1}^{\rm i}=N_{N_1}^{\rm eq} = 3/4$. At $K\simeq 3$ a rapid
transition takes place from strong to weak washout. Even here analytical
and numerical results agree within 30\%. For the case of zero initial $N_1$ 
abundance one obtains an interpolation formula $\k_{\rm f}(K)$ analogous
to Eq. (\ref{kfan}).

\begin{figure}[t]
\centerline{\psfig{file=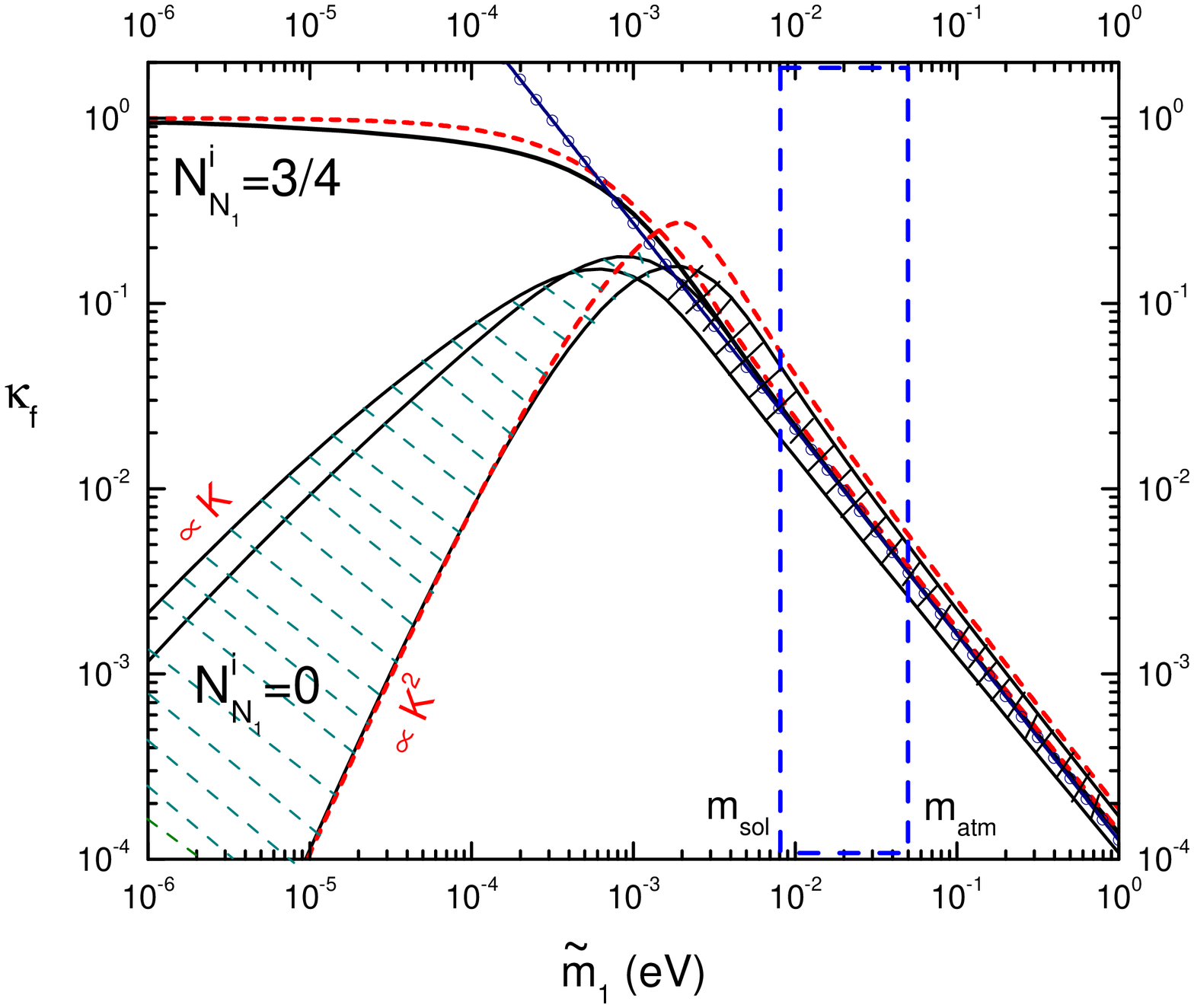,height=8cm,width=12cm}}
\caption{\small Final efficiency factor when the washout term $\Delta W$
is neglected. From \cite{bdp04}.}
\label{KFIN}
\end{figure}

The above discussion of decays and inverse decays can be extended to
include $\Delta L=1$ and $\Delta L=2$ scattering and washout processes.
In the weak washout regime, $K\ll 1$, the main effect is that the efficiency 
factor of Eq. (\ref{reduction}) is enhanced to $\k_{\rm f} \propto K$.
Relevant effects include scattering processes
involving gauge bosons \cite{pu04,gnx04} and thermal
corrections to the decay and scattering rates \cite{gnx04,crx98}.
The range of different results is represented in Fig. (\ref{KFIN}) by
the hatched region. An additional uncertainty in the weak washout
regime is due to the dependence of the final results on the
initial $N_1$ abundance and a possible initial asymmetry
created before the onset of Leptogenesis.

The situation is very different in the strong washout regime. Here the
final efficiency factor is not sensitive to the neutrino production
because a thermal neutrino distribution is always reached at high
temperatures. For $\mt > m_* \simeq 10^{-3}\,{\rm eV}$,  the effect of 
$\Delta L=1$ processes on the washout is not larger than about $50\%$,
as indicated by the hatched region in Fig. (\ref{KFIN}). Within these uncertainties,
the final efficiency factor is given by the simple power law:
\begin{equation}\label{plaw}
  \k_{\rm f}=(2\pm1)\,10^{-2}\,
             \left({0.01\,{\rm eV}\over\mt}\right)^{1.1\pm 0.1}\;.
\end{equation}
Both the scale of solar neutrino oscillations, $m_{\rm sol}
\equiv \sqrt{\Delta m^2_{\rm sol}}\simeq 8\times 10^{-3}\,{\rm eV}$, and the
scale of atmospheric neutrino oscillations, $m_{\rm atm} \equiv
\sqrt{\Delta m^2_{\rm atm}} \simeq 0.05\,{\rm eV}$, are larger
than the equilibrium neutrino mass $m_*$. Hence, the range of neutrino masses,
and therefore $\mt$, indicated by neutrino oscillations 
lies entirely in the strong washout regime where
theoretical uncertainties are small and the efficiency factor is still
large enough to allow for successful Leptogenesis.

\subsection{Bounds on Neutrino Masses}

The $\D L=2$ processes with heavy neutrino exchange generate a contribution
to the washout rate that depends on the absolute neutrino mass scale 
$\mb^2 = m_1^2 + m_2^2 + m_3^2$,
\begin{equation}
\D W \, \propto \,  {M_{\rm P}M_1\, \mb^2\over v_F^4}  \; .
\end{equation}
As long as $\D W$ can be neglected, the efficiency factor is independent 
of $M_1$. With increasing $\mb$ however, the washout rate $\D W$ becomes 
important and eventually prevents successful Leptogenesis. This leads to the 
upper bound on the absolute neutrino mass scale. \cite{bdp02,bdp022}

\begin{figure}[t]
\centerline{\psfig{file=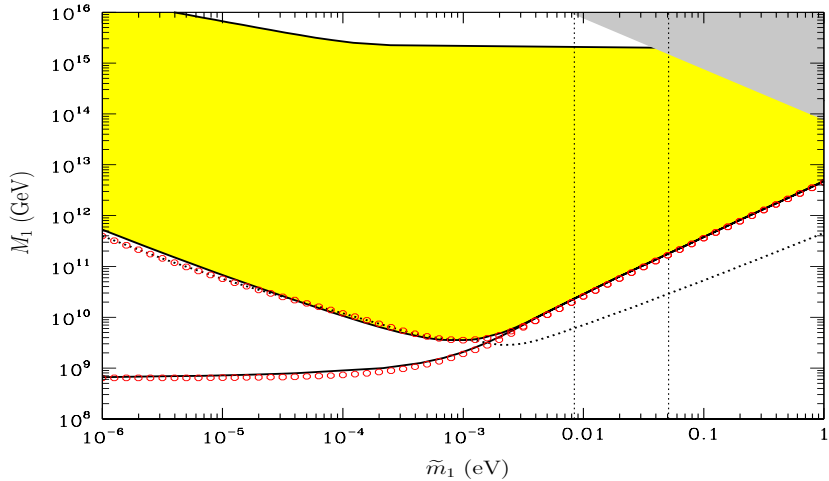,height=6.5cm,width=11cm}}
\caption{Analytical lower bounds on $M_1$ (circles) and $T_{\rm i}$ 
(dotted line) for $m_1 = 0$, $\eta_B^{CMB} = 6\times 10^{-10}$ and 
$m_{\rm atm} = 0.05\,{\rm eV}$.
The analytical results for $M_1$ are compared with the numerical ones 
(solid lines). Upper and lower curves correspond to zero and thermal
initial $N_1$ abundance, respectively.
The vertical dashed lines indicate the range ($m_{\rm sol}$,$m_{\rm atm}$).
The gray triangle at large $M_1$ and large $\mt$ is excluded by theoretical
consistency. From \cite{bdp04}.}
\label{BTR}
\end{figure}

One can also obtain a lower bound on the heavy neutrino masses \cite{di}, 
because
the CP asymmetry $\ve_1$ satisfies an upper bound \cite{di,hmy,bdp2,hlx},
which is a function of $M_1$, $\mt$ and $\mb$. Since the rates entering the
Boltzmann equations depend on the same quantities, there exists for arbitrary
neutrino mass matrices a maximal baryon asymmetry $\eta^{\rm max}_{B}$, 
\begin{eqnarray}
\eta_{B} &\leq& \eta^{\rm max}_{B}(\mt,M_1,\mb) \NO\\
&\simeq& 0.01\ \ve_1^{\rm max}(\mt,M_1,\mb)\ \k(\mt,M_1\mb^2)\;.
\end{eqnarray}
Requiring the maximal baryon asymmetry to be larger than the observed one, 
\begin{equation}\label{cmbcon}
\eta^{\rm max}_{B}(\mt,M_1,\mb) \geq  \eta^{CMB}_{B} \;,
\end{equation}
then yields a constraint on the neutrino mass parameters $\mt$, $M_1$ 
and $\mb$. For each value of $\mb$ one obtains a domain in the 
($\mt$-$M_1$)-plane, which 
is allowed by successful baryogenesis. For $\mb \geq 0.20$~eV 
this domain shrinks to zero. One can easily translate this 
bound into upper limits on the individual neutrino masses. In a similar way,
one finds a lower bound on $M_1$, the smallest mass of the heavy Majorana 
neutrinos. The resulting upper and lower bounds are \cite{bdp2}
\begin{equation}
m_i < 0.1\,{\rm eV}\;, \quad M_1 > 4 \times 10^8~\mbox{GeV}\;,
\end{equation}
where we have assumed thermal initial $N_1$ abundance. The upper bound on 
the light neutrino masses holds for a normal as well as an inverted hierarchy of masses. 
For zero initial $N_1$ abundance one obtains the more restrictive lower
bound $M_1 > 2 \times 10^9~\mbox{GeV}$. For $\mt > m_*$, the baryon asymmetry 
is generated at the temperature $T_B \simeq M_1/z_B < M_1$. Hence the lower 
bound on the reheating temperature $T_i$ is less restrictive than the lower
bound on $M_1$. The results of a detailed
analytical and numerical calculation are summarized in Fig. (\ref{BTR}).
For the lower bound on the reheating temperature one finds 
$T_i > 2 \times 10^9\ {\rm GeV}$ \cite{bdp04,gnx04}.\footnote{In the 
supersymmetric case the CP asymmetry is enhanced but also 
the washout processes are stronger. These two effects partly compensate each 
other \cite{plu98}, leading to the slightly less stringent bound
$T_i > 1.5 \times 10^9\ {\rm GeV}$ \cite{db04}.} 

What is the theoretical error on the upper bound for the light neutrino
masses? In order to answer this question one needs a full quantum 
mechanical treatment of Leptogenesis, a 
challenging problem! A possible starting point is the Kadanoff-Baym 
equations for which a systematic expansion around the Boltzmann equations
can be constructed \cite{bf00}. One then has to calculate relativistic 
corrections, off-shell effects, `memory effects', higher order loop 
corrections, etc.
One important effect is the running of neutrino masses between the Fermi
scale and the energy scale of Leptogenesis \cite{bcx00,akx03}. Also relevant 
are
the $\D L=1$ scattering processes involving gauge bosons \cite{pu04,gnx04}.
Conceptually interesting are thermal corrections at large
temperatures, $T > M_1$, which correspond to loop corrections involving
gauge bosons and the top quark \cite{gnx04}. Their effect is large if thermal 
masses are treated as kinematical masses in the evaluation of scattering 
matrix elements. At sufficiently high temperatures the process
$N_1 \rightarrow H \ell_L$ is then kinematically forbidden whereas the process
$H \rightarrow N_1 \ell_L^{\dagger}$ is allowed by `phase space'. On the
contrary, thermal correction are small if they are only included as
propagator effects \cite{crx98}. It is important to clarify this issue 
for the treatment of non-equilibrium processes at high temperatures.

The analysis \cite{gnx04} leads to the upper bound on the light
neutrino masses $m_i < 0.15$~eV. In \cite{bdp04} an upper bound 
of $0.12$~eV has 
been obtained. About $0.02~{\rm eV}$ of this difference is due to
the different treatment of radiative corrections \cite{akx03}, the remaining
$0.01~{\rm eV}$ reflects differences in the treatment of thermal corrections. This discrepancy has to be compared with an uncertainty of
about $-0.02~{\rm eV}$ due to `spectator processes' \cite{bp}, which have
not been taken into account in both analyses. Hence, within the minimal
seesaw model and the present status of theoretical calculations, the upper 
bound on the light Majorana neutrino masses is now known rather precisely.

The main result of this section is summarized in Fig. (\ref{KFIN}).
For $\mt > m_*$, the efficiency factor, and therefore
the baryon asymmetry $\eta_B$, is independent of the initial $N_1$ abundance. 
Furthermore, the final baryon asymmetry does not depend on the value of an 
initial baryon asymmetry generated by some other mechanism \cite{bdp2}. Hence,
the value of $\eta_B$ is entirely determined by neutrino properties. 
In this way Leptogenesis singles out the neutrino mass range 
\begin{equation}
10^{-3}~{\rm eV} < m_i < 0.1~{\rm eV}\;.
\end{equation}
The firm predictions of thermal Leptogenesis open a window into the physics
of the early universe at temperatures $T_B = {\cal O}(10^{10}~ {\rm GeV})$,
and we can ask what the implications are for dark matter, cosmology and
particle physics.

\subsection{Triplet Models and Resonant Leptogenesis}

Measurements in neutrino physics determine the parameters of the neutrino mass 
matrix,
\begin{equation}
m_\n = -m_D^T{1\over M}m_D + m_\n^{\rm triplet}\; ,
\end{equation}
which in general contains a contribution from $SU(2)$ triplet fields \cite{wet}
in addition to the seesaw term generated by $SU(2)$ singlet heavy Majorana 
neutrinos. So far, we have only considered the minimal case, 
$m_\n^{\rm triplet}=0$. Clearly, a dominant triplet 
contribution would destroy the connection between Leptogenesis and low energy 
neutrino physics.

The discovery of quasi-degenerate neutrinos with masses above the bound 
$0.1~{\rm eV}$ would require significant modifications of minimal Leptogenesis
and/or the seesaw mechanism. In this case $SU(2)$ triplet contributions to 
neutrino masses could be a possible way out \cite{tri,hlx,ak}. Clearly,
one then has no upper bound on the light neutrino masses anymore. 
Yet Leptogenesis with right-handed neutrino decays can still work
yielding a slightly relaxed lower bound on the heavy neutrino masses. 
For instance, one may have $m_i \sim 0.35~{\rm eV}$ with 
$M_1 > 4 \times 10^8~{\rm GeV}$ \cite{ak}.

Another way to reconcile quasi-degenerate light neutrinos with Leptogenesis
makes use of the enhancement of the CP asymmetry in case of 
quasi-degenerate heavy neutrinos \cite{fps}. For instance, to raise the
upper bound from $0.1~{\rm eV}$ to $0.4~{\rm eV}$, a degeneracy of 
$\D M/M$ for the heavy neutrinos in the range $0.4 - 10^{-3}$ is required, 
depending on assumptions about the neutrino mass matrices \cite{bdp2,hlx}. 
In the extreme case of `resonant Leptogenesis' \cite{pu04}, CP asymmetries 
$\ve = {\cal O}(1)$ are reached for degeneracies 
$\D M/M = {\cal O}(10^{-10})$. 
In this case the right-handed neutrino masses may be as small as 
$1~{\rm TeV}$, which may lead to observable signatures at
colliders. A number of models of this type have been constructed \cite{rlm},
some of which make use of the relative smallness of soft supersymmetry
breaking terms \cite{rls}.

\subsection[Connection with Low Energy CP Violation]{Is the CP Violation in Leptogenesis Connected with the Low 
Energy CP Violation in the Neutrino Sector?}

As was shown in Section 3,  the seesaw model has 6 CP-violating phases in
the Yukawa matrix $h_{ij}$. Leptogenesis depends on one combination
of these 6 phases. However, there are only 3 CP-violating phases at low 
energies. Hence it is impossible to determine all 6 phases in the
theory, even if one were to measure all 3 low-energy phases. 
Futhermore, as we discussed earlier, one of these low-energy phases remains undetermined by
experiments feasible at low energies.

Nevertheless, the effective number of high energy
CP-violating phases is reduced if one of the superheavy Majorana
neutrinos $N_i$ is extremely heavy and decouples from the seesaw system. 
In this case, the Yukawa couplings $h_{ij}$ effectively are given by a $2\times 3$ 
matrix that contains 6 complex parameters and hence 6 phases. Three of
the 6 phases can be absorbed into the wave functions of $\ell_L$ and
thus one is left with only 3 CP-violating phases at high energies. In this case,  the 3 low-energy 
CP-violating phases that appear in the neutrino mass matrix $m_\nu$ are reduced to
only two physical phases, because ${\rm det}(m_\nu)\simeq
0$. Although these 2 low-energy phases are, in principle, measurable in future experiments, this is still
not enough to determine all three phases in the full
theory. Therefore, even in this simplified example, one cannot establish any link between the sign of the
Universe's baryon-number asymmetry with the observable CP-violating
phases at low energies.

In the very special case where $h_{ij}$ has two zeros, one has
only one CP-violating phase. In this case the CP-violating phase in
neutrino oscillations is connected  with the phase in 
Leptogenesis or, equivalently,  the sign of the baryon-number asymmetry in
the Universe \cite{FGY}.\footnote{The prediction of the sign of 
the CP-violating phase in neutrino oscillations depends on which heavy
Majorana neutrino is responsible for Leptogenesis. This problem is
solved in the inflaton-decay scenario in supersymmetry (SUSY) theories, 
because one choice
is unable to produce enough lepton-number asymmetry due to the
constraint on the reheating temperature $T_R < 10^7$ GeV \cite{KKM}.}
Thus, in this case, one may indeed test directly the idea of Leptogenesis.
It is interesting that such a restricted model, where  $h_{13}=h_{21}=0$, 
is still consistent with data on neutrino oscillations.

\section{Dark Matter Considerations}

It is certainly possible that the mechanism that generates a primordial
matter-antimatter asymmetry in the Universe is not physically related to
the existence of a non-luminous component of the energy density of the
Universe, a component that now accounts for about 25 $\%$ of the total
energy density. However, it would be very interesting if these two
phenomena, so central to the history of the Universe, were connected in
some deep way. It turns out, as we shall see, that if Leptogenesis is
the mechanism by which a primordial matter-antimatter asymmetry in the
Universe is established, it considerably impacts what the dark matter in
the Universe can be.  

Of the three viable options for dark matter, from the point of view of
particle physics, two are either linked or constrained by thermal
Leptogenesis and the third has clear connections to nonthermal
Leptogenesis. Before discussing these points in some detail, it is
useful to briefly review the extant dark matter candidates motivated by
particle physics.

\subsection{PQ Symmetry and Axions}

It is well known that QCD admits the presence of an additional CP violating 
term in its Lagrangian density \cite{tH},
\begin{equation}
{\cal{L}}_{\theta}= \frac{\alpha_3 }{8 \pi}\theta F^{\mu \nu}_i
\tilde{F}_{i \mu \nu},
\end{equation}
where $\tilde{F}_{i\mu\nu}=1/2\epsilon_{\alpha\beta\mu\nu}F^{\alpha\beta}_i$.
If $\theta$ is non-vanishing ${\cal{L}}_{\theta}$, which is C-even and
P-odd,  violates CP and T invariance. Because possible CP
violating parameters of the strong interactions, like the electric
dipole moment of the neutron, are very tightly bounded by experiment
\cite{edm}, the parameter $\theta$ must be very small ($\theta \leq
10^{-10}$) \cite{bound}. The reason for this is a mystery, and is known as
the Strong CP Problem \cite{RDPCP}.

Probably the most `natural' solution suggested for the Strong CP Problem
is to assume that the total Lagrangian for the strong and electroweak
interactions is invariant under a global chiral $U(1)_{PQ}$ symmetry
\cite{PQ}. Even though this symmetry is spontaneously broken, one can
show \cite{PQ} that as a result of the $U(1)_{PQ}$ symmetry the parameter
$\theta$ is driven to zero. In effect, what happens is that the 
CP violating Lagrangian term  ${\cal{L}}_{\theta}$ is replaced by a 
CP conserving interaction between the CP odd pseudo-Goldstone 
boson\footnote{Axions are not true Goldstone bosons because the $U(1)_{PQ}$
symmetry is anomalous \cite{WW}. In fact, the same effective potential
for axions that serves to drive $\theta$ to zero gives axions a small
mass. This mass is of order \cite{RDPCP} $m_ a^2 \sim m_q
\Lambda_{QCD}^3/f_a^2$,
where $f_a$ is the scale where the $U(1)_{PQ}$ symmetry breaks down
spontaneously, and $m_q$ is the (light) quark mass.} associated with the
spontaneous breakdown of $U(1)_{PQ}$ -- the axion -- \cite{WW} and
$F\tilde{F}$:
\begin{equation}
{\cal{L}}_{\theta} \to \frac{\alpha_3 }{8 \pi }\frac{a}{f_a} F^{\mu \nu}_i
\tilde{F}_{i \mu \nu}.
\end{equation}

Originally, it was supposed  \cite{PQ} that the $U(1)_{PQ}$ symmetry was
broken at the electroweak scale. Then $f_a \sim v_F \simeq 200$ GeV and
the axion mass lies in the keV range. However, axions in this mass
range, which are coupled with strength $1/f_a$, have been ruled out by
experiment \cite{RDPCP}. Astrophysical considerations, however, impose
very strong  constraints on axions much lighter than a keV, as their
emission from stars would significantly alter their properties. Only
axions that are sufficiently weakly coupled (hence, with large enough
$f_a$ and thus a correspondingly small mass) avoid these constraints,
and one finds the bound \cite{Raffelt} $f_a  \geq 10^{10}$ GeV.  On the
other hand, $f_a$ cannot be arbitrarily large, because zero-momentum
axion oscillations in the early Universe would carry enough energy
density (proportional, approximately, to $f_a$) to overclose the
Universe \cite{Cbounds}. Thus, for an appropriate value for $f_a$,
axions can be the dark matter in the Universe. In particular, one finds
\cite{Sikivie} that $f_a \simeq 10^{12}$ GeV gives $\Omega_a \simeq 1$.

\subsection{ Dark Matter Candidates from Supersymmetry}

Supersymmetry, a boson-fermion symmetry, has been invoked extensively as
the solution of the so-called hierarchy problem. This problem is
related to the fact that without some stabilizing influence radiative
corrections in the electroweak theory would naturally push the Fermi
scale $v_F$ to have the value of whatever cutoff delimits the validity
of the theory. Typically, this cutoff is imagined to be at the Planck
scale $M_P$, and why $v_F << M_P$ is the hierarchy problem. This problem
is resolved if there is some low energy (spontaneously broken)
supersymmetry. Due to the fermion-boson nature of supersymmetry,
radiative corrections of parameters in the electroweak theory (like $v_F$) are now
only logarithmically dependent on the cutoff, not quadratically
dependent. Hence, effectively, if there is some low energy supersymmetry
one can contemplate having a hierarchy like $v_F<<M_P$, because radiative
shifts can only change $v_F$ logarithmically.

In general, supersymmetric theories possess a discrete symmetry
(R-symmetry) that distinguishes particles from their supersymmetric
partners. As a result, the lightest supersymmetric particle (the LSP) is
stable and, in principle, could be the source of the dark matter in the
Universe. Indeed, it is known \cite{KT} that  the energy density
of particles of mass of $O(v_F)$, whose annihilation cross section is of
electroweak strength, is of the order of the critical energy density
that closes the Universe. With supersymmetric partners of ordinary
particles having electroweak scale masses and interactions, the LSP is
therefore an ideal candidate for the dark matter in the Universe
\cite{SDM}. In this review we will discuss both the cases of neutralinos (the
SUSY partners of gauge and Higgs boson) and of gravitinos (the spin 3/2
partner of the graviton) as LSP candidates.

\subsection{Extended Structures}

Scalar fields are necessary ingredients of the standard electroweak
model, as well as its supersymmetric extension. It is well known that
theories with scalar fields can lead to the formation of nontopological
solitons. These extended structures, known as Q-balls \cite{Coleman},
may be stable or unstable and arise when some scalar field carries a
conserved $U(1)$ charge. For example, in supersymmetric theories
sleptons and squarks carry, respectively, lepton and baryon number.

In supersymmetric theories, more generally, Q-balls can develop along
flat directions of the scalar potential \cite{DK}. These Q-balls can, in
a number of instances, carry baryon number. If the baryon number of the
Q-ball is large enough, and its mass is small enough, the baryonic
Q-balls are stable. Because of their stability, one can imagine that
these Q-balls could be the dark matter in the Universe.\footnote{This, 
however, is not easily achieved because, in general, the squarks are 
unstable with their baryon number eventually residing on quarks. If the 
squarks are light enough, stability can be achieved. However, as Kasuya et al.
point out \cite{KKT}, it is difficult to explain both the baryon asymmetry 
and the dark matter density simultaneously. Nevertheless, there are scenarios 
where unstable Q-balls are the source for both baryogenesis and neutralino 
dark matter \cite{FY02}.} Typically \cite{DK}, if stable Q-balls exist they have both very large baryon number ($B \sim 10^{26}$)
and are very massive ($M_Q \leq 10^{26}$
GeV). Unfortunately, this makes their detection very difficult, because
their flux is very low \cite{Ara}.

 \subsection{ Natural Connection of Axions with Leptogenesis}

The scale of $ U(1)_{PQ}$ breaking needed for axions to be the dark
matter in the Universe ($f_a \simeq 0.3 \times 10^{12}$ GeV) is close
enough to the mass of the lightest right handed neutrino ($M_1 \simeq
10^{10}$ GeV) needed for Leptogenesis to seek for a common linkage. In
fact, the existence of such a linkage was observed long ago by
Langacker, Peccei and Yanagida \cite{LPY}. What these authors observed
was that if $M_1$ were due to the VEV of a scalar field $\sigma$, one
could identify this field as carrying a PQ-symmetry rather that lepton
number. 

Let us examine this assertion in a bit more detail by looking at the
Yukawa interactions of the quarks and leptons with the three Higgs
fields\footnote{For a PQ symmetry to exist one needs to have two 
$SU(2)$ doublet Higgs fields, $\phi_1$ and $\phi_2$, rather than just the 
single Higgs field of the Standard Model $H$ (and its Hermitian adjoint $H^{\dagger}$).}  $\phi_1$, $\phi_2$ and $\sigma$.
Schematically, one has
\begin{equation} \label{LPQ}
{\cal{L}}_{\rm{Yukawa}}= h_{\sigma} \sigma N_RN_R + h \bar{N}_R
\phi_2 \ell_L +  h_u \bar{u}_R \phi_2 q_L +  f_d \bar{d}_R \phi_1
q_L +  f \bar{e}_R \phi_1 {\ell}_L + {\rm h.c.}\, .
\end{equation}
One sees that Eq. (\ref{LPQ}) is invariant under a PQ-symmetry, where
\begin{eqnarray}
\phi_1, \phi_2&\to& e^{i \alpha} \phi_1, e^{i \alpha} \phi_2\\
N_R, {\it{l}}_R, u_R, d_R&\to& e^{i \alpha} N_R, e^{i \alpha}
{\it{l}}_R, e^{i \alpha}  u_R, e^{i \alpha} d_R, 
\end{eqnarray}
provided that
\begin{equation}
\sigma \to e^{-2i \alpha} \sigma.
\end{equation}

To allow $<\sigma>=f_a  >>v_F$, as in all invisible axion models
\cite{DFSZ,KSVZ}, requires one fine tuning. In the above case,
this requires the PQ-invariant term in the scalar potential
\begin{equation}
V=\kappa \sigma \phi_1 \phi_2 + {\rm h.c.} 
\end{equation} 
to have the constant $\kappa \sim v_F^2/f_a$, to allow electroweak
symmetry breaking to occur at a scale much below the scale of $U(1)_{PQ}$
symmetry breaking ($ v_F << f_a$). 

The Yukawa couplings of
this model guarantee that the mass of the lightest right-handed neutrino
and $f_a$ are related: $M_1= 2(h_{\sigma})_{11}f_a$. Thus, if axions are the source of the dark matter energy density in the Universe and the baryon aymmetry arises from Leptogenesis, because $\Omega_{\rm{DM}} \sim f_a$ and $\Omega_{\rm{B}} \sim M_1 \sim f_a$, their ratio is independent of the scale of $U_{PQ}(1)$ breaking. Hence it is perhaps not surprising that this ratio is of order unity.

\subsection{The Gravitino Problem in Supersymmetric Theories}

As we discussed in Section 4, for Leptogenesis to be effective, the mass
of the lightest right-handed neutrino has to be greater than 
$2\times 10^9$ GeV. This bound, in turn, means that thermal Leptogenesis must
have occurred at temperatures above $2 \times 10^{9}$ GeV. Hence, if the
Universe went through an inflationary period, as all evidence seems to
suggest \cite{WMAP}, the reheating temperature after inflation $T_R$
must have been greater than $2 \times 10^{9}$ GeV for Leptogenesis to be
the source of the matter-antimatter asymmetry in the Universe. This high
reheating temperature is problematic for supersymmetric theories because
it leads to an overproduction of light states, like the gravitino, with
catastrophic consequences for the evolution of the Universe after
inflation. Unless these observational inconsistencies can be avoided, it
appears that Leptogenesis in supersymmetric theories cannot produce the
desired baryon asymmetry in the Universe.

The production of gravitinos after inflation has been studied in some
detail \cite{KL}. The thermal production of gravitinos
produced by the strong interactions of quarks, squarks, gluons and
gluinos is governed by the Boltzmann equation \cite{BBB}
\begin{equation} \label{evn3}
\frac{dn_{3/2}}{dt} +3Hn_{3/2}=C_{3/2}(T),
\end{equation}
where
\begin{equation}
C_{3/2}(T)=\frac{3\zeta(3)\alpha_3(T)}{\pi^2}
\frac{T^6}{M_P^2}\left(1+\frac{m_{\tilde{g}}^2}{3m_{3/2}^2}\right)F(T).
\end{equation}
Here $F(T)$ is a thermal factor of O(10) and $m_{\tilde{g}}$ and
$m_{3/2}$ are, respectively, the gluino and the gravitino masses.
Integrating Eq. (\ref{evn3}) to a reheating temperature $T_R$, the resulting
relic density of produced gravitinos is given by
\begin{equation}
\Omega_{3/2}h^2\simeq 0.44 \alpha_3(T_R) \left[1
+\frac{1}{3}\left(\frac{\alpha_3(T_R)}{\alpha_3(\mu)}\right)^2
\left(\frac{m_{\tilde{g}}(\mu)}{m_{3/2}}\right)^2\right]
\left(\frac{T_R}{10^{10}\rm{GeV}}\right)
\left(\frac{m_{3/2}}{\rm{100GeV}}\right),
\end{equation}
where $h$ is the scaled Hubble parameter and $\mu \sim M_Z$.

 If gravitinos are stable (i.e. they are the LSP), the WMAP constraint on
the amount of dark matter in the Universe \cite{WMAP}
\begin{equation}
\Omega_{DM}h^2= 0.1126^{+ 0.0161}_{-0.0181}
\end{equation}
constrains $\Omega_{3/2}h^2$ to be below this value and, for any given
reheating temperature $T_R$ and gravitino mass $m_{3/2}$, gives a bound
on the gluino mass. If, on the other  hand, the gravitinos are not
stable, their rate of production for $T_R>2 \times 10^{9}$ GeV is so
large that subsequent gravitino decays completely alter the standard Big
Bang Nucleosynthesis (BBN) scenario. Thus, in either case, there are
severe constraints imposed on supersymmetric dark matter, which we will
discuss in detail below.

\begin{figure}[t]
\centerline{\psfig{file=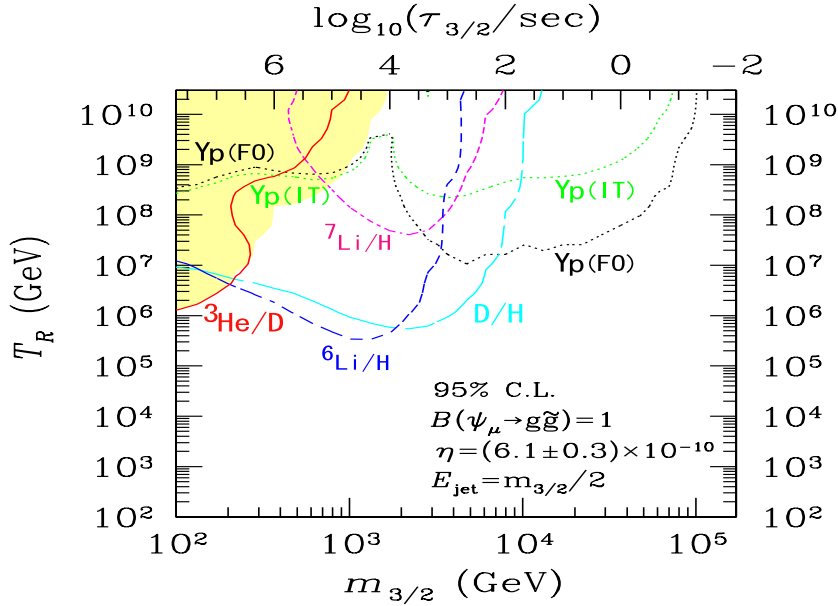,height=8cm,width=11cm}}
\caption{Upper bounds on the reheating temperature as function of the 
gravitino mass for the case where the gravitino dominantly decays into a
gluon-gluino pair. From \cite{KKM}.}
\label{KKMfig}
\end{figure}

If the gravitino is unstable, it has a long lifetime and decays during 
or after BBN for an interesting range of 
the gravitino mass, $m_{3/2}\simeq 100 ~{\rm GeV}-10 ~{\rm TeV}$. 
The gravitino decay products destroy the light elements produced by the BBN 
and 
hence the relic abundance of gravitinos is constrained from above to keep the
success of the BBN \cite{Falom}. This leads to an upper bound of the reheating temperature 
$T_R$ after inflation, since the abundance of gravitinos is proportional to
the reheating temperature.  A recent detailed analysis derived a strigent 
upper bound $T_R< 10^{6-7}$ GeV when the gravitino decay has hadronic  
modes (see Fig. (\ref{KKMfig})) \cite{KKM}. 
This upper bound is much lower than the temperature for 
Leptogenesis, $T_R > 2\times 10^{9}$ GeV \cite{bdp04,gnx04}. 
Therefore,  thermal Leptogenesis seems difficult to reconcile with low energy supersymmetry if gravitino masses lie in the range $m_{3/2}\simeq 100~{\rm
GeV}-10$ TeV - a natural range for Supergravity (SUGRA) models.

\subsection[Solutions to the Gravitino Problem in Thermal Leptogenesis]{Solutions to the Gravitino Problem in Thermal\\ Leptogenesis}

There have been several attempts to solve the gravitino
problem in thermal Leptogenesis. Here we will briefly review a number of these proposed solutions.

One possibility has been proposed by Pilaftsis who considers quasi-degenerate 
heavy Majorana neutrinos $(M_1\simeq M_2)$ \cite{enhancement}. In this model 
the lepton-asymmetry parameter $\ve$ is enhanced by a factor of $M_1/(M_1-M_2)$ 
and hence the decays of both $N_1$ and $N_2$ may produce enough asymmetry even for 
$T_R < 10^{6-7}$ GeV. However, it is difficult to find a compelling justification for having such a degeneracy in the heavy neutrino spectrum.

Another proposal was made by Bolz, Buchm\"uller and Pl\"umacher \cite{BBP}, who
consider the case where the gravitino is the stable lightest SUSY particle 
(LSP). In this case the next lightest supersymmetric particle (NLSP)
is the subject of the cosmological constraint, because 
its decay products may destroy the light elements created by the BBN, much like
the unstable gravitino. Detailed analyses show that this scenario 
favors the $\tilde{\tau}$ NLSP compared to the neutralino NLSP and gravitino
masses below $m_{3/2} \simeq 100\ {\rm GeV}$ \cite{FIY,RRA}.\footnote{The 
gravitino mass is even less cnstrained if the LSP is a scalar
neutrino or the gluino. For a class of supergravity models an upper bound of
$5\times 10^9$ GeV on the reheating temperature has been obtained \cite{RRA}.}
In general, the
gravitino production can be dominated by NLSP decays \cite{frt03} or by
thermal processes \cite{bhr03}.

A third proposed solution makes use of gauge-mediation model of supersymmetry breaking in which the gravitino
is the stable LSP  with a mass $m_{3/2} < 1$ GeV. It turns out that, if the 
gravitino mass is $m_{3/2} < 16$ eV \cite{viel}, 
then there is no gravitino problem. However, in this case the gravitino cannot be the dark matter. It must be something else, perhaps the axion. For the range of gravitino masses
 $m_{3/2}\simeq 100 ~{\rm keV}-1~ {\rm GeV}$, there is the interesting 
possibility
that late-time entropy production in a class of gauge mediation models can  
naturally make the gravitino the dominant component of dark matter \cite{FIY2}.
In this scenario the reheating temperature can be as high as 
$T_R\simeq 10^{13}$
GeV. A light axino with mass ${\cal O}(1\ {\rm keV})$ as LSP 
and a gravitino as NLSP would solve the gravitino problem \cite{ay00}.

Finally, another possible solution arises if supersymmetry breaking effects
are transmitted to the Standard Model sector through the scale anomaly,
resulting  in very heavy gravitino masses $m_{3/2}>100$ TeV. 
In this case the gravitino decays before the time of BBN and hence there is no 
cosmological problem. However, the gravitino decay modes contain always one LSP and 
hence the relic abundance of the gravitino must be constrained from above so 
that 
the density of the nonthermal LSP produced by the gravitino decays does not exceed 
the dark-matter density. 
This condition leads to $T_R< 10^{11}$ GeV \cite{GGW,IKMY}, which is consistent
with thermal Leptogenesis.

 We stress here that each of the above `solutions' predicts distinct particle spectra at the
TeV scale, which 
may be testable in future collider experiments at the 
Large Hadron Collider (LHC). 
If one discovers supersymmetry but all the above possibilities are  
excluded experimentally, this would argue strongly against thermal 
Leptogenesis although nonthermal Leptogenesis could still be viable.  
On the other hand, if some of these scenarios are confirmed experimentally, thermal Leptogenesis will become much more compelling.

\subsection{Leptogenesis and Lepton Flavor Violation in SUSY Models}

Lepton flavor violation is another area in which thermal Leptogenesis and supersymmetry may have  some linkages. If the neutrinos have a mass, lepton flavor violation (LFV) processes such as
$\mu\rightarrow e + \gamma$ decay can occur. In non supersymmetric theories, these processes are strongly 
suppressed by a factor of $(m_\nu /M_W)^2$ in rate  
and hence are unmeasurable  physically. However, this is not the case in the SUSY 
Standard Model, if the seesaw mechanism is effective.

In the SUSY Standard Model scalar quarks and leptons are assumed to have
a universal SUSY-breaking soft mass, $m_0$, at the Planck scale. 
Otherwise one would have too large flavor-changing neutral currents (FCNC).
However, even then quantum corrections resulting from Yukawa interactions of the quarks  
generate a violation of the universality of the soft masses for scalar
quarks, which induces FCNC.  
In the lepton sector the Yukawa couplings of the superheavy
Majorana neutrinos $N_i$ also generates non-universal masses 
for scalar leptons that serves as a source of LFV \cite{BM}.\footnote{The 
Yukawa couplings $h_{ij}$ of the Higgs to the $N_i$s and leptons induce flavor 
dependent soft masses for scalar leptons.
At the one-loop level the induced mass is given by
 $ (m_{\tilde \ell})^2_{ij} \simeq -6m^2_0/(4\pi)^2h^{\dagger}_{ik}h_{kj}
ln(M_P/M_k)$,
where $m_0$ is the universal soft mass for scalar leptons at the  Planck
scale $M_P$. Thus, one may obtain information on the high-energy Yukawa coupling
$h_{ij}$ and the  heavy neutrino masses $M_{k}$ by measuring directly the mass matrix for the scalar 
leptons. However, the phases in $ h^{\dagger}h$ are different from the phases 
contributing to Leptogenesis \cite{di01}.}  If the
relevant Yukawa couplings are of $O(1)$, or equivalently $M_3\simeq 10^{15}$
GeV, one  predicts a branching ratio for  $\mu\rightarrow e + \gamma$ decay 
that may be
testable in future experiments. However, an accurate prediction for LFV
processes is very difficult, since it hinges on unkown Yukawa coupling 
constants \cite{MS}. In particular, the constraint coming from Leptogenesis 
that $M_1\simeq 10^{10}$ GeV is not strong enough to suggest that LFV 
processes have potentially testable rates.

\section{Nonthermal Leptogenesis}

Supersymmetry is an important symmetry for the unification 
of all interactions and all matter, and the SUSY Standard Model is
considered as a plausible scenario for producing new physics at the TeV scale. Thus, it is quite 
interesting to consider theories where supersymmetry is spontaneously broken in a hidden sector connected to ordinary matter by gravitational strength interactions-- the SUGRA framework.  The seesaw 
mechanism is easily incorporated into this framework. However, as we discussed in some detail in Section 5, the gravitino problem argues against thermal Leptogenesis, particularly in SUGRA.

A possible solution to this problem may be provided by nonthermal Leptogenesis 
\cite{AD, Shafi,KMY, MSYY,MrYa,MrYa2}, 
where one does not have a strong constraint on the reheating temperature. We 
will discuss here specifically  nonthermal Leptogenesis via inflaton decay  
\cite{Shafi,KMY}, which we consider an interesting scenario. 
In the next subsection, we present general arguments for this
scenario and show that it suggests a lower bound on the mass of the 
heaviest light neutrino $m_3 > 0.01$ eV.
In the subsequent subsection, we will also discuss the Affleck-Dine mechanism
\cite{AD} for Leptogenesis which, specifically in supersymmetric theories, 
 is also an interesting 
mechanism to generate the matter-antimatter asymmetry.
 
\subsection{Nonthermal Leptogenesis via Inflaton Decay}

Inflation early on in the history of the Universe is one of the most attractive hypothesis in modern 
cosmology, because it not only solves long-standing problems in cosmology, 
like the horizon and the flatness problems \cite{guth}, but also 
accounts for the origin of density fluctuations \cite{density-fluctuations}.
In this subsection we discuss the hypothesis that the inflaton $\Phi$ decays
dominantly into a pair of the lightest heavy Majorana neutrinos,  
$\Phi\rightarrow N_1+N_1$. We assume, for simplicity, that other decay modes
including those into pairs of $N_2$ and $N_3$ are energetically forbidden. 
The produced $N_1$ neutrinos decay subsequently into $H + \ell_L$ 
or ${H}^{\dagger} + \ell_L^{\dagger}$. If the reheating temperature $T_R$ is lower
than the mass $M_1$ of the heavy neutrino $N_1$, then the out-of-equilibrium 
condition \cite{Sakharov} is automatically satisfied. 

The above two channels for $N_1$ decay have different branching ratios 
when CP conservation is violated. Interference between tree-level and 
one-loop diagrams generates a lepton-number asymmetry \cite{l-asymmetry}.
Following our discussion in Section 4, the lepton asymmetry parameter 
$\ve$ can be  written as \cite{MrYa,MrYa2,susycp} 
\footnote{Because of supersymmetry,
the asymmetry parameter $\ve$ below is a factor of 2 larger than that 
given in Eq. (\ref{CPas}).}
\begin{equation}
\ve = -\frac {3}{8\pi}\frac{M_1}{\langle H\rangle ^2}
           m_{3}\delta _{\rm eff}\,,
\end{equation}
where the effective CP-violating phase $\delta_{\rm eff}$ is given by
\begin{equation}
\delta_{\rm eff} = \frac{ {\rm Im}\left[h^2_{13} + \frac{m_{2}}{m_{3}}
h^2_{12} + \frac{m_{1}}{m_{3}}h^2_{11}\right]}{|h_{13}|^2 + |h_{12}|^2
+ |h_{11}|^2}\;.
\end{equation}
Numerically, one obtains for the $\ve$ 
parameter 
\begin{equation}
\ve \simeq -2\times 10^{-6} \left( \frac{M_1}{10^{10}{\rm GeV}}\right)
\left(\frac{m_{3}}{0.05 {\rm eV}}\right)\delta_{\rm eff}\;.
\end{equation}

The chain decays  $\Phi\rightarrow N_1 + N_1$ and $N_1\rightarrow H + \ell_L$ 
or $H^{\dagger} + \ell_L^{\dagger}$ reheat the Universe producing not only the 
lepton-number asymmetry but also entropy for the thermal bath.
The ratio of the lepton number to entropy density after reheating 
is estimated to be \cite{KMY}
\begin{eqnarray}
\frac{n_L}{s} &\simeq& -\frac{3}{2}\ve\frac{T_R}{m_{\Phi}} \nonumber \\
   &\simeq& 3\times 10^{-10} \left(\frac{T_R}{10^6{\rm GeV}}\right)
\left(\frac{M_1}{m_\Phi}\right)
      \left(\frac{m_{3}}{0.05{\rm eV}}\right),
\end{eqnarray}
where $m_{\Phi}$ is the inflaton mass and we have taken $\delta_{\rm eff} =1$.
This lepton-number asymmetry is converted into a baryon-number asymmetry 
through the sphaleron effects and one obtains \cite{ht}
\begin{equation}
\frac{n_B}{s} \simeq -\frac {8}{23}\frac {n_L}{s}\;.
\end{equation} 

We should stress, here, an important merit of the inflaton-decay
scenario: It does not require a reheating temperature $T_R \sim M_1$,
but it requires only $m_{\Phi} > 2M_1$. On the other hand, for thermal 
Leptogenesis to work it is necessary that $T_R \sim M_1$, 
which necessitates higher reheating temperature for Leptogenesis to 
produce enough matter-antimatter asymmetry.

If one assumes that $T_R < 10^7$ GeV to satisfy the cosmological constraint on 
the gravitino abundance \cite{KKM} discussed earlier and
uses $m_\Phi > 2M_1$,  the observed baryon number to entropy ratio \cite{WMAP}
gives a constraint on the heaviest light neutrino:
\begin{equation} \label{m3}
m_{3} > 0.01~ {\rm eV}.
\end{equation}
It is very interesting 
that the neutrino mass suggested by atmospheric neutrino oscillation experiments, 
$\sqrt{\Delta m^2_{\rm atm}}\simeq 0.05$ eV, just satisfies the above
constraint. However, to get this bound 
 we assumed that the inflaton decays dominantly
into a pair of $N_1$s. If this branching ratio is only 10 $\%$, the
lower bound on the neutrino mass exceeds the observed neutrino mass
$\sqrt{\Delta m^2_{\rm atm}}\simeq 0.05$ eV.

A variety of models have been considered to restore the bound of 
Eq. (\ref{m3}) by imposing a symmetry. However, it is perhaps most 
interesting to consider that the
scalar partner of the heavy Majorana neutrino $N_1$ is the inflaton
itself \cite{MSYY},  and the inflaton decay into a lepton plus a Higgs boson gives an effective branching ration of 100\%.
In this model, one must assume that the initial
value of the scalar partner of $N_1$ is much larger than the Planck
scale to cause inflation (chaotic inflation \cite{linde}).
However, chaotic inflation is not easily realized in SUGRA, because 
the minimal supergravity potential has 
an exponential factor, ${\rm exp}(\phi^*\phi /M^2_G)$, that prevents 
any scalar field $\phi$ from having a value larger than the reduced Planck scale $M_G \simeq 2.4\times 10^{18}$ GeV. Ref. \cite{GL} uses a restricted form of
the Kahler potential.

\subsection{Affleck-Dine Leptogenesis}

In the SUSY Standard Model, in the limit of unbroken supersymmetry, some combinations of scalar fields do not enter
the potential, constituting so-called flat directions 
of the potential.
Since the
potential is almost independent of these fields, they may have large initial values in the early Universe.  Such flat directions 
receive soft masses in the SUSY-breaking vacuum.  When the expansion rate
$H_{\rm exp}$ of the Universe becomes comparable to their masses, the flat directions begin to
oscillate around the minimum of the potential. If the
flat directions are made of scalar quarks and carry baryon number, the
baryon-number asymmetry can be created
during these coherent oscillations. This is the
Affleck-Dine (AD) mechanism for Baryogenesis \cite{AD}.

QCD corrections, however, make the potential of the AD fields milder 
than $|\phi|^2$. This allows non-topological soliton solutions (Q-balls) 
\cite{Q-ball2} to form in the early Universe, as a result of the coherent oscillations in the flat
directions. Because Q-balls
have long lifetimes, their decays produce a huge  amount of entropy at late times.  To avoid this problem one must choose parameters in the SUSY theory so that the density
of the lightest SUSY particle (LSP) does not exceed the dark matter
density in the present Universe \cite{Q-ball2}. Although this may not be a problem, it is much safer to consider flat directions without QCD
interactions, because such directions most likely do not have Q-ball solutions. 

The most interesting candidate \cite{MrYa} for such a flat direction is
\begin{equation}
\phi_i = (2H\ell_i)^{1/2},
\end{equation}
where $\ell_i$ is the lepton doublet field of the $i$-$th$ family. 
Here, $H$ and $\ell_i$ represent the scalar components of the corresponding 
chiral multiplets. The Yukawa
interactions of $H$ make the potential of $\phi_i$ steeper than
the mass term and hence there is no instability of the coherent oscillation
(i.e. there are no Q-ball solutions). Because this flat direction carries
lepton number, a lepton asymmetry will be created during the coherent
oscillation (AD Leptogenesis) \cite{MrYa}. Sphaleron processes then transmute,
 in the usual fashion, this lepton asymmetry into a baryon asymmetry. 

The seesaw mechanism induces a dimension-five operator in the
superpotential for the theory,\footnote{For ease of notation we have dropped the subscript $i$ below.}
\begin{equation}
W=\frac{m_{\nu}}{2|\langle H\rangle |^2}(\ell H)^2,
\end{equation}
where we have used a basis in which the neutrino mass matrix is
diagonal. With this superpotential we have a SUSY-invariant potential
 for the flat direction $\phi$ given by
\begin{equation}
V_{\rm SUSY}=\frac{m_{\nu}^2}{4|\langle H\rangle |^4}|\phi|^6.
\end{equation}
In addition to the SUSY-invariant potential we have a SUSY-breaking
potential,
\begin{equation}
\delta V= m_{\phi}^2|\phi|^2 + \frac{m_{\rm SUSY}m_{\nu}}{8|\langle H\rangle |^2}
(a_m\phi^4 + {\rm h.c.}).
\end{equation}
Here, $a_m$ is a complex number. We take 
$m_{\phi} \simeq m_{\rm SUSY} \simeq 1$ TeV and $|a_m|\sim 1$.
The second term in $\delta V$ is very important, because it gives rise to
the lepton-number generation.

We assume that the flat direction $\phi$ acquires a negative $({\rm
mass})^2$  induced by the inflaton potential and rolls down to the point 
balanced by the SUSY-invariant potential $V_{\rm SUSY}$ during 
inflation. Thus, the AD field $\phi$ has an initial value of 
$\sqrt{H_{\rm inf}|\langle H\rangle |^2/m_{\nu}}$, where $H_{\rm inf}$ 
is the Hubble constant (the expansion rate) during inflation. $\phi$ decreases in
amplitude gradually after inflation, and begins to oscillate around
the potential minimum when the Hubble constant $H_{\rm exp}$ of the
Universe becomes
comparable to the SUSY-breaking mass $m_{\phi}$. 
At the beginning of the oscillation, the AD field
has the value $ |\phi_{0}| \simeq \sqrt{m_{\phi}|\langle H\rangle
|^2/m_{\nu}}$ which, as shown below, is an effective initial value for 
Leptogenesis. 

Let us consider now lepton-number generation in this scenario. The evolution of the AD
field $\phi$  is described by
\begin{equation}
\frac{\partial^2\phi}{\partial t^2} + 3H_{\rm exp}\frac{\partial \phi}{\partial t} 
+ \frac{\partial V}{\partial \phi^*} = 0 \;,
\end{equation}
where $V=V_{\rm SUSY} + \delta V$.
Because the lepton number is given by 
\begin{equation}
n_{\rm L} = i\left(\frac{\partial \phi^*}{\partial t}\phi - 
\phi^*\frac{\partial \phi}{\partial t}\right)\;,
\end{equation}
 the evolution of $n_{\rm L}$ is given by
\begin{equation}
\frac{\partial n_{\rm L}}{\partial t} + 3H_{\rm exp}n_{\rm L} 
= \frac{m_{\rm SUSY}m_{\nu}}{2|\langle H\rangle |^2}
{\rm Im}(a_m^*\phi^{*4})\;.
\end{equation}

The motion of $\phi$ in the phase direction generates the lepton number. This is predominantly created just after the AD field $\phi$
starts its coherent oscillation, at a time $t_{\rm osc} \simeq 1/H_{\rm
osc}\simeq 1/m_{\phi}$, because the amplitude $|\phi|$ damps as $t^{-1}$
during the oscillation. Thus, we obtain for the lepton number 
\begin{equation}
n_{\rm L}\simeq \frac{m_{\rm SUSY}m_{\nu}}{2|\langle 
H\rangle |^2}\delta_{\rm eff}|a_m\phi_0^4|\times t_{\rm osc}\;,
\end{equation}
where $\delta_{\rm eff} = {\rm sin}(4{\rm arg}\phi +{\rm arg}a_m)$
represents an effective CP-violating phase. Using $m_{\rm SUSY}\simeq
m_\phi$, $ |\phi_0| \simeq \sqrt{m_{\phi}|\langle H\rangle
|^2/m_{\nu}}$ and $t_{\rm osc}\simeq 1/m_\phi$, we find
\begin{equation}
n_{\rm L}\simeq \delta_{\rm eff}m_{\phi}^2 \frac{|\langle 
H\rangle |^2}{2m_{\nu}}\;.
\end{equation}

After the end of inflation, the inflaton begins to oscillate around
the potential minimum and $n_{\rm L}/\rho_{\rm inf}$
stays constant until the inflaton decays. Here $\rho_{\rm inf}$ 
is the energy density of the inflaton. The inflaton decay reheats the
Universe producing entropy $s$. Because $\rho /s = 2T_R/4$, we find for the lepton-number asymmetry the expression
\begin{equation}
\frac{n_{\rm L}}{s} \simeq \left( \frac{\rho_{\rm inf}}{s}\right)
\left(\frac{n_{\rm L}}{\rho_{\rm inf}}
\right) \simeq \delta_{\rm eff}\frac{3T_R}{4M_{ G}}
\frac{|\langle H\rangle |^2}{6m_{\nu}M_{G}}\;.
\end{equation}
Here we have used $\rho_{\rm inf} \simeq 3m_{\phi}^2M_{G}^2$ at the
begining of the AD field oscillation (when most of the lepton number 
is generated). This lepton-number asymmetry is converted to a
baryon-number asymmetry by the KRS mechanism. In this way one obtains for the baryon-number asymmetry
\begin{equation}
\frac{n_{\rm B}}{s} \simeq \frac{1}{23}
\frac{|\langle H\rangle |^2T_R}{m_{\nu}M_{G}^2}\;.
\end{equation}
The observed ratio $n_{\rm B}/s \simeq 0.9\times 10^{-10}$ implies 
$m_\nu \simeq 10^{-9}$ eV for $T_R \simeq 10^6$ GeV. This 
small mass corresponds to the mass of the lightest neutrino. We should 
note that for such a low reheating temperature one may neglect the 
effects due to thermal mass for the AD field $\phi$ \cite{FHY}. 

\section{Conclusions and Summary of Results}

In this article we have discussed the physical mechanism responsible for the origin of matter in the Universe. Both the rather
large observed value for the ratio of baryons to photons, $\eta_B$, in the present epoch and the absence of antimatter are the consequences of a primordial asymmetry between matter and antimatter generated early on in the Universe. Although a variety of mechanisms have been proposed for producing this primordial asymmetry, in this review we have focused on Leptogenesis as the origin of matter. In our view, this is the most appealing scenario for the origin of matter, for at least three reasons:

1) Explicit lepton number violation is very natural once one includes right-handed neutrinos in the Standard Model. Furthermore, the lightness of the observed neutrinos strongly suggests, through the seesaw mechanism, the presence of superheavy neutrinos, whose decays can produce a lepton-antilepton asymmetry.

2) Quantum mechanically, through the KRS mechanism, one can automatically 
turn a leptonic asymmetry into a baryonic asymmetry. Indeed, because of the 
existence of these sphaleron processes, the origin of matter is linked to 
phenomena in the early Universe that result in the establishment of a 
(B-L)-asymmetry, like Leptogenesis.

3) If neutrino masses lie in the range $ 10^{-3} \rm{eV} < m_i< 0.1 \rm{eV}$, as suggested by neutrino oscillation experiments, the leptonic asymmetry produced in thermal Leptogenesis is both independent of the abundance of heavy neutrinos and of any pre-existing asymmetry and has the right magnitude to yield the observed value for $\eta_B$.

Because, in the final analysis, $ \eta_B$ is just one number, it is important 
to ask if the particular mechanism proposed for the origin of matter has 
other consequences. Thus in this review we examined in some detail how 
Leptogenesis fit with ideas proposed to explain the dark matter that 
constitutes about 25\% of the Universe's energy density.\footnote{We did not 
try to examine models of dark energy in the light of Leptogenesis, because our understanding of dark energy is still in its infancy.} We pointed out that axionic dark matter is perfectly compatible with Leptogenesis. Indeed, it is possible to very naturally link the scale of the heavy neutrinos with that of $U(1)_{PQ}$ breaking $f_a$, so that the ratio $\Omega_{\rm DM}/ \Omega_{B}$ is independent of these large scales. The situation, however, is more complex in the case of supersymmetric dark matter.

To be effective, thermal Leptogenesis needs to occur at high temperatures, above $T= 2 \times 10^9$ GeV. This means that the Universe after inflation must have reheated to at least this temperature. However, in supersymmetric theories such a high reheating temperature is problematic as it leads to an overproduction of gravitinos. When they decay,
gravitinos of such abundances completely alter the primordial abundance of light elements produced in Big Bang Nucleosynthesis.
The gravitino problem, however,  is not fatal as there are a number of ways to mitigate the overproduction of gravitinos. Nevertheless, if Leptogenesis is at the root of the origin of matter, the supersymmetric spectrum at low energies and the nature of the LSP are quite constrained. Thus, in a sense, Leptogenesis is also quite predictive in this context.

Although much of our review, very naturally, focused on thermal Leptogenesis, 
we also discussed two examples where matter originated  through a leptonic 
asymmetry produced in nonthermal processes. These models, although much more 
speculative, illustrate  some of the possible other options for the origin of 
matter. Naturally, in this case some of the specific predictivity is lost.

\vspace{0.5cm}
\noindent
{\bf Acknowledgments}\\
\noindent
In our work on the topics discussed in this review we have benefitted from
the insight of many colleagues. W.B. and T.Y. owe special thanks to
M. Bolz, A. Brandenburg, P. Di Bari, K. Hamaguchi, M. Ibe, K. Izawa, 
T. Moroi, M. Pl\"umacher and M. Ratz. The work of T.Y. has been supported
in part by a Humboldt Research Award. RDP's work was supported in part by the 
Department of Energy under Contract No. FG03-91ER40662, Task C.


\end{document}